\documentclass[prc,aps,longbibliography,superscriptaddress,nofootinbib]{revtex4-2}
\usepackage{graphicx}
\usepackage{amssymb}
\usepackage{amsmath}
\usepackage{dcolumn}
\usepackage{color}
\usepackage{times}
\usepackage{txfonts}
\usepackage[colorlinks=true,citecolor=blue,linkcolor=blue,urlcolor=blue]{hyperref}
\usepackage{multirow}
\usepackage{longtable}
\DeclareSymbolFont{EulerExtension}{U}{euex}{m}{n}
\DeclareMathSymbol{\euintop}{\mathop} {EulerExtension}{"52}
\DeclareMathSymbol{\euointop}{\mathop} {EulerExtension}{"48}

\begin{document}
\title{
EXFOR-based simultaneous evaluation of neutron-induced uranium and plutonium fission cross sections for JENDL-5
}
\author{Naohiko Otuka}
\email{n.otsuka@iaea.org}
\affiliation{
Nuclear Data Section,
Division of Physical and Chemical Sciences,
Department of Nuclear Sciences and Applications,
International Atomic Energy Agency,
A-1400 Wien, Austria
}
\affiliation{
Nishina Center for Accelerator-Based Science,
RIKEN,
Wako-shi, Saitama 351-0198,  Japan;\\
}
\author{Osaumu Iwamoto}
\email{iwamoto.osamu@jaea.go.jp}
\affiliation{
Nuclear Data Center,
Nuclear Science and Engineering Center,
Japan Atomic Energy Agency,
Tokai-mura, Naka-gun, Ibaraki 319-1195, Japan
}
\date{\today}
\begin{abstract}
The neutron-induced fission cross sections were simultaneously evaluated for the JENDL-5 library for $^{233,235}$U and $^{239,241}$Pu from 10~keV to 200~MeV and for $^{238}$U and $^{240}$Pu from 100~keV to 200~MeV.
%
%
Evaluation was performed by least-squares fitting of Schmittroth's roof function to the logarithms of the experimental cross sections and cross section ratios in the EXFOR library.
A simultaneous evaluation code SOK was used with its extension to data in arbitrary unit.
The outputs of the code were adopted as the evaluated cross sections without any further corrections.
The newly obtained evaluated cross sections were compared with the evaluated cross sections in the JENDL-4.0 library and the IAEA Neutron Data Standards 2017.
The evaluated cross sections were also validated against the californium-252 spontaneous fission neutron spectrum averaged cross sections, $\Sigma\Sigma$ (coupled thermal/fast uranium and boron carbide spherical assembly) neutron spectrum averaged cross sections, and small-sized LANL fast system criticalities.
The changes in the obtained evaluated cross sections from those in the JENDL-4.0 library are within 4\% ($^{241}$Pu), 3\% ($^{233}$U, $^{240}$Pu), or 2\% ($^{235}$U, $^{239}$Pu).
The newly evaluated $^{235}$U, $^{238}$U and $^{239}$Pu cross sections agree with the IAEA Neutron Data Standards 2017 within 2\% with some exceptions.

(This is a preprint of N.~Otuka and O.~Iwamoto, J. Nucl. Sci. Technol., to be published \href{http://doi.org/10.1080/00223131.2022.2030259}{online} as an open access article.)
\end{abstract}
\maketitle
\newpage
\section{Introduction}
Fission is a major reaction channel in the interaction of low-energy neutrons with various actinide nuclides.
Nuclear reaction systems such as thermal and fast reactors undergo fission,
and the fission cross sections must be accurately known to understand the behavior of the systems.
Fission cross sections such as $^{235,238}$U(n,f) cross sections are also important due to their use as references (standards)~\cite{IAEA2007International,Carlson2009International,Carlson2018Evaluation}
for determination of neutron fluence.
The combination of the Hauser-Feshbach formalism and Hill-Wheeler formula describes fission cross sections for fast neutrons fairly well,
and it has been implemented in various nuclear reaction model codes~\cite{Iwamoto2007Development,Kawano2021CoH3,Koning2012Modern,Herman2007EMPIRE}.
Prediction of fission cross sections by this approach is sensitive to choice of parameters such as barrier heights,
which must be adjusted to reproduce the experimental fission cross sections~\cite{Iwamoto2007Development}.
On the other hand, fission events are relatively easy to detect,
and many experimental datasets are available in the EXFOR library~\cite{Otuka2014Towards}.
For production of a nuclear data library, therefore,
it is adequate to evaluate fission cross sections by fitting to the experimental cross sections without modelling the physics,
and to evaluate the cross sections of other channels by a reaction model code with fission parameters adjusted to reproduce the evaluated fission cross sections~\cite{Iwamoto2016CCONE}.

Experimental fission cross sections for fast neutrons are often determined with the aid of reference cross sections.
The measured ratio of the fission cross section to the reference cross section is free from choice of the reference cross section,
and the ratio has been routinely compiled in the EXFOR library without conversion to the absolute cross section.
One can include the cross section ratio to evaluation after conversion to the absolute cross sections by using a state-of-the-art reference cross section.
But a more sophisticated way is to treat absolute cross sections (e.g., $^{239}$Pu and $^{235}$U fission cross sections) and their ratio (e.g., $^{239}$Pu/$^{235}$U fission cross section ratio) in an equivalent manner without converting one to the other.
Sowerby et al.~\cite{Sowerby1970Simultaneous,Sowerby1974Simultaneous} and Poenitz~\cite{Poenitz1970Interpretation} are known as pioneers in this approach,
and the JENDL project also has adopted this technique since its early versions.
In its first attempt for the JENDL-2 library,
evaluation of the $^{235,238}$U and $^{239,240,241}$Pu(n,f) cross sections from 100~eV to 20~MeV was divided into three steps 
--- (1)~preliminary evaluation for $^{235}$U, (2)~normalization and evaluation for other target nuclides by using the preliminary $^{235}$U evaluation result, and (3)~reevaluation for $^{235}$U to meet the requirements from the second step ---
to maintain consistency between the cross sections of all nuclides~\cite{Matsunobu1979Simultaneous,Matsunobu1980Simultaneous}.
%
%
This first attempt was altered in JENDL-3 evaluation by developing a new simultaneous evaluation procedure based on least-squares fitting of the logarithms of the cross sections and their ratios to a B-spline function~\cite{Uenohara1983Simultaneous,Uenohara1983Significance,Kikuchi1992Japanese}.
This method was applied to evaluation of $^{235,238}$U, $^{239,240,241}$Pu(n,f) and $^{197}$Au, $^{238}$U(n,$\gamma$) cross sections from 50~keV to 20~MeV~\cite{Kanda1986Simultaneous} for the JENDL-3 library~\cite{Shibata1990Japanese},
which cross sections were also subsequently adopted in the JENDL-3.2 library with modifications such as replacement of the $^{235}$U(n,f) cross section above 13~MeV~\cite{Nakagawa1995Japanese}.

For the JENDL-3.3 library~\cite{Shibata2002Japanese},
a new simultaneous evaluation code SOK (Simultaneous evaluation on KALMAN) was developed for reevaluation of the $^{233,235,238}$U and $^{239,240,241}$Pu fission cross sections~\cite{Kawano2000Simultaneous,Kawano2000Evaluation}.
The JENDL-3.3 cross sections were further updated with SOK,
and the updated cross sections were adopted in JENDL Actinoid File 2008 (JENDL/AC-2008)~\cite{Iwamoto2009JENDL} and JENDL-4.0 library~\cite{Shibata2011Japanese}.
Apart from the JENDL project,
the SOK code also analyzed the experimental database prepared for the IAEA Coordinated Research Project (CRP) ``International Evaluation of Neutron Cross-Section Standards",
and the Project concluded that SOK and another least-squares analysis code GMA~\cite{Poenitz1997Simultaneous} give very similar results or distinctive differences that can be readily explained~\cite{IAEA2007International,Carlson2009International}.

As the evaluated data libraries tend to include more covariance information,
the Nuclear Reaction Data Centres (NRDC) start to pay more attention to compilation of experimental uncertainty and covariance information,
and also have extended the EXFOR format by introducing correlation property flags and computer readable format for matrices~\cite{Otuka2012Experimental,Smith2012Experimental,Zerkin2013Extension,Otuka2014Documentation}.
Meanwhile, a number of new experimental datasets relevant to the simultaneous evaluation were obtained at the time-of-flight facilities such as CERN n\_TOF and LANSCE.
The data newly published by CERN n\_TOF collaboration have been compiled in the EXFOR library under tight collaboration with OECD NEA Data Bank and IAEA Nuclear Data Section~\cite{Dupont2017Dissemination}.
From the view of validation,
there was a progress in compilation of benchmark field neutron spectra and cross sections measured in the benchmark fields for the IRDFF-II project~\cite{Trkov2020IRDFF-II}.

Considering these movements,
we decided to perform simultaneous evaluation of $^{233,235,238}$U and $^{239,240,241}$Pu fission cross sections again by construction of an experimental database dedicated to the new evaluation.

This article reports the procedures and results of the new simultaneous evaluation for the JENDL-5 library~\cite{Iwamoto2020Status}.
We will compare the newly evaluated cross sections with the evaluated cross sections in the JENDL-4.0 library and the IAEA Neutron Data Standards 2017,
and also discuss validation of the newly evaluated cross sections against the californium-252 prompt fission neutron spectrum averaged cross sections, $\Sigma\Sigma$ (coupled thermal/fast uranium and boron carbide spherical assembly) spectrum averaged cross sections, and the small-sized LANL fast system criticalities.

\section{Cross sections in major general-purpose and standard libraries}
\label{sec:majorlibs}
The fission cross sections of the six target nuclides treated in the present work are also included in the latest versions of major national general-purpose libraries (e.g., BROND-3.1~\cite{Blokhin2016New}, CENDL-3.2~\cite{Ge2020CENDL}, ENDF/B-VIII.0~\cite{Brown2018ENDF}, JEFF-3.3~\cite{Plompen2020Joint} and JENDL-4.0).
Table~\ref{tab:libraries} summarizes the origins of the evaluated cross sections to the best of our knowledge based on the description in the ENDF files (MF1 MT451) and evaluation summary documents (e.g., Ref.~\cite{Tang1991Evaluation,Cai1992Evaluation,Liang1992Evaluation} for the CENDL-3.2 cross sections taken from CENDL-2).
This table shows that the evaluations are usually based on experimental works (data) but sometimes also done by statistical model calculation, which is also a reasonable choice especially for the subthreshold fission regions of $^{238}$U and $^{240}$Pu (below $\sim$1~MeV).

We also observe that these general purpose libraries often adopt the cross sections evaluated for the IAEA Neutron Data Standards (IAEA-2006~\cite{IAEA2007International,Carlson2009International} and IAEA-2017~\cite{Carlson2018Evaluation}) ENDF/B-VI Neutron Cross Section Standards~\cite{Carlson1993ENDF}.
The cross sections in these standard libraries are obtained by GMA and R-matrix analysis.
Their evaluations include not only $^{235,238}$U and $^{239}$Pu fission cross sections but also other important standard reactions such as $^6$Li(n,t)$^4$He, $^{10}$B(n,$\alpha$)$^7$Li and $^{197}$Au(n,$\gamma$)$^{198}$Au cross sections.
The thermal neutron constants (cross sections and average total neutron multiplicities) of $^{233}$U, $^{235}$U, $^{239}$Pu and $^{241}$Pu as well as the spontaneous fission neutron multiplicity of $^{252}$Cf have been also evaluated from the very early stage of this IAEA activity~\cite{Westcott1965Survey,Hanna1969Revision,Lemmel1975ThirdIAEAEvaluation,Lemmel1975ThirdIAEAReview},
and they are included in the experimental database of the GMA analysis (GMA database) since some measured cross sections are normalized to those at the thermal energy.
The n-p scattering is also evaluated separately, and the experimental data of other standard reactions are renormalized in terms of the newly evaluated n-p scattering before GMA and R-matrix fitting.
Namely, the experimental cross sections included in the IAEA standard evaluations are ultimately normalized with the latest values of the most fundamental n-p scattering cross section.
Our present evaluation also uses the fission cross sections measured relative to the $^6$Li(n,t)$^4$He and $^{10}$B(n,$\alpha$)$^6$Li standard cross sections, but our experimental database adopts the measured ratios converted by the experimentalists to the absolute cross sections with the $^6$Li and $^{10}$B standard cross sections at the time, and therefore our evaluation may be more biased by the old standard cross sections.

Two other major differences in the SOK and GMA evaluations are (1) treatment of cross section ratio data, and (2) selection of the energy nodes.
\begin{itemize}
\item SOK linearizes the cross section ratio by logarithmic transformation while GMA linearizes the cross section by the first-order Taylor approximation (c.f. Eq.~13b of Ref.~\cite{Poenitz1981Data}).
\item
GMA requires extrapolation of the originally reported experimental cross sections to an energy grid commonly chosen for all reactions (c.f. Fig.~4 of Ref.~\cite{Poenitz1981Data}) but SOK performs fitting to Schmittroth's roof function which does not require the extrapolation neither common energy grid structure.
\end{itemize}
The GMA database has some spectrum averaged cross sections and they are utilized in fitting in evaluation for the IAEA Neutron Data Standards while the current evaluation uses spectrum averaged cross sections only for validations.
It would be also worthwhile to mention that the GMA database often includes the numbers not reported by the experimentalist but estimated by the evaluators while our experimental database constructed for the SOK analysis is automatic conversion of the EXFOR library and solely based on the numbers provided by the authors except for (1) assignment of the correlation properties (e.g., uncorrelated or fully correlated) and (2) addition of a few constant uncertainties and correlations not explicitly written in the source articles and cannot be compiled in their EXFOR entriesbecause of the EXFOR policy.
\begin{table}
\caption{Evaluation procedures of $^{233,235,238}$U and $^{239,240,241}$Pu fast neutron fission cross sections in recent general purpose libraries.}
\label{tab:libraries}
\begin{center}
\centering\includegraphics[clip,angle=0,width=\linewidth]{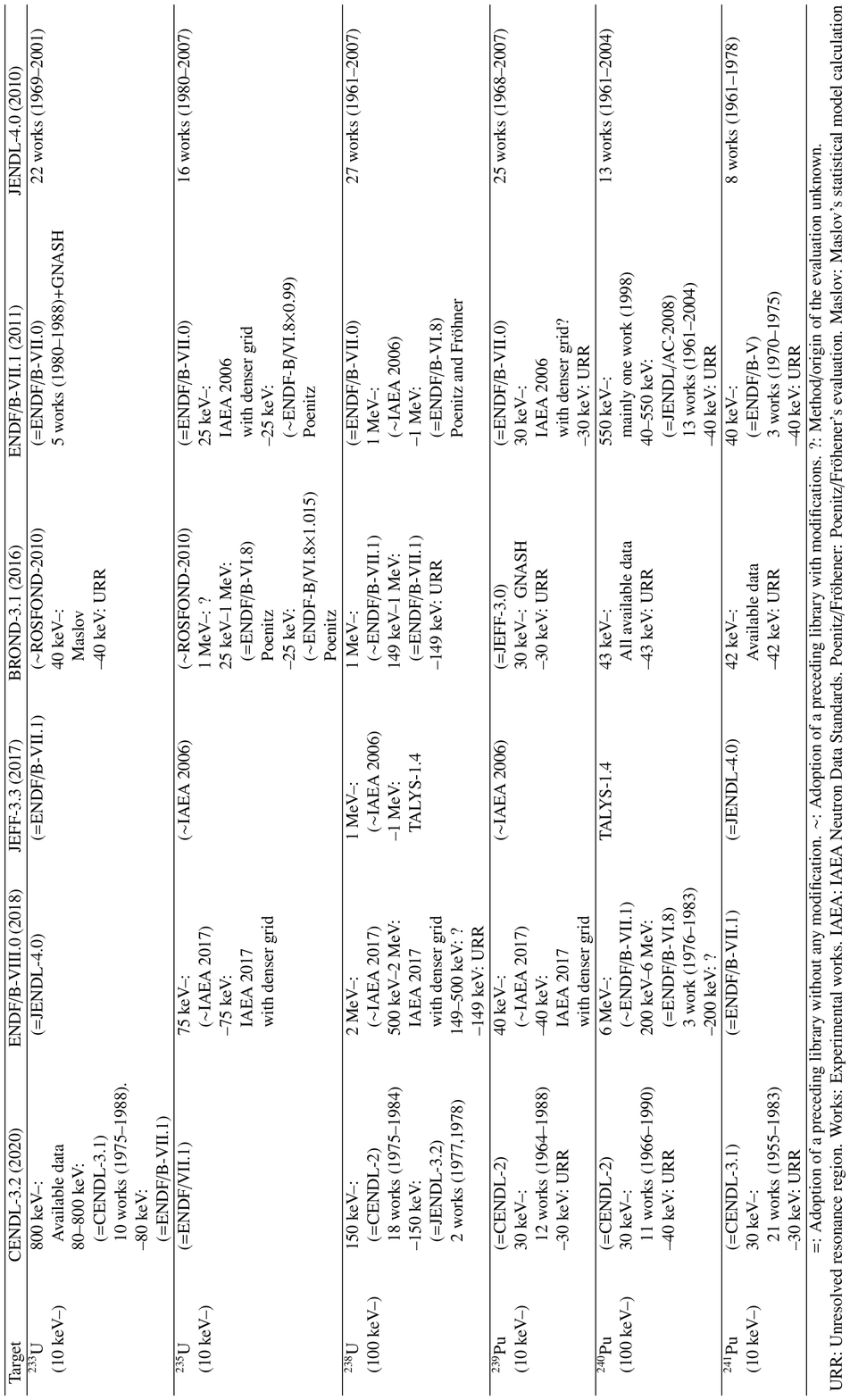}
\end{center}
\end{table}

\section{Experimental database}
The experimental database is the main input to the present least-squares analysis.
All experimental datasets
\cite{
Calviani2009High,
Tovesson2004233Pa,
Guber2000New,
Shpakov1989Absolute,
Zasadny1984Measurement,
Dushin1983bStatistical,
Murzin1980Measurement,
Zhagrov1980Fission,
Poenitz1978Absolute,
Gwin1976Measurement,
Yan1975Measurement,
Blons1971MeasurementAnalysis,
Tovesson2014Fast,
Belloni2011Neutron,
Shcherbakov2002Neutron,
Shpak1998Angular,
Meadows1988Fission,
Manabe1988Measurements,
Kanda1986Measurement,
Carlson1978aMeasurement,
Fursov1978Measurement,
Meadows1974Ratio,
Pfletschinger1970Measurement,
Amaducci2019Measurement,
Nolte2007Cross,
Carlson1992Measurements,
Lisowski1991Fission,
Merla1991Absolute,
Kalinin1991Correction,
Iwasaki1988Measurement,
Li1988Absolute,
Carlson1985Absolute,
Dias1985Application,
Weston1984Subthreshold,
Li1983Absolute,
Wasson1982Absolute,
Wasson1982Measurement,
Mahdavi1983Measurements,
Cance1981Measures,
Arlt1981aAbsoluteFission,
Salvador-Castineira2017Absolute,
Miller2015Measurement,
Meadows1996Measurement,
Eismont1996Relative,
Meadows1989Search,
Wu1983Measurement,
Hu1980Measurement,
Alkhazov1979Fission,
Cance1978Absolute,
Vorotnikov1976Sub,
Behrens1976Measurements,
Wen2020Measurement,
Casperson2018Measurement,
Paradela2015High,
Li1989Ratio,
Kanda1985Measurement,
Garlea1984Measuring,
Goverdovskii1984aMeasurement,
Varnagy1982New,
Difilippo1978Measurement,
Behrens1977Measurements,
Fursov1977aMeasurement,
Cierjacks1976Measurements,
Nordborg1976Fission,
Cance1976Measurements,
Meadows1975Ratio,
Poenitz1972Measurements,
Meadows1972Ratio,
Weston1992High,
Zhou1983Fast,
Ryabov1979Measurements,
Davis1978Absolute,
Szabo1976Measurement,
Blons1970MesureHaute,
Szabo1973Mesure,
Schomberg1970Ratio,
Szabo1970New,
Tovesson2010Cross,
Staples1998Neutron,
Weston1983Neutron,
Meadows1978Fission,
Kari1978Measurement,
Fursov1977bMeasurement,
Gayther1975Measurement,
Poenitz1972Additional,
Szabo1971235U,
Poenitz1970Measurement,
Salvador-Castineira2015Neutron,
Gul1986Measurements,
Aleksandrov1983Neutron,
Cance1983Mesures,
Budtz-Jorgensen1981Neutron,
Khan1980New,
Stamatopoulos2020Investigation,
Laptev1998Fast,
Iwasaki1990Measurement,
Behrens1983Measurement,
Meadows1981Fission,
Wisshak1979Measurement,
Kupriyanov1979Measurement,
Behrens1978Measurements,
Frehaut1974Mesures,
Wagemans1982241Pu,
Carlson1977Measurement,
Blons1970MesureAnalyse,
Blons1971MeasurementFission,
Kaeppeler1973Measurement}
were taken from the EXFOR library without any changes unless otherwise noted in this section.
When the uncertainty information written in the source article is missing in the EXFOR entry,
we notified it to the Data Centre maintaining the EXFOR entry,
and the Centre updated the EXFOR entry accordingly.
Table~\ref{tab:expsummary} summarizes the number of datasets used for final fitting.
%
%
A summary of all experimental datasets adopted in the present evaluation (e.g., EXFOR subentry number, first author, year of publication, laboratory, reference) is published elsewhere~\cite{Otuka2022Input}.
\footnote{
This summary overwrites the list of the EXFOR datasets in MF1 MT451 of the JENDL-5 library ENDF-6 file.
The experimental database in an ASCII file is available as a Supplemental Material.
}.
\begin{table}
\caption{Number of experimental datasets of fission cross sections and their ratios for final fitting.}
\label{tab:expsummary}
\begin{center}
\begin{tabular}{cccccccc}
\hline
            &\multicolumn{6}{c}{Numerator}\\
\cline{2-7}
Denominator & $^{233}$U  & $^{235}$U & $^{238}$U & $^{239}$Pu & $^{240}$Pu & $^{241}$Pu \\
\hline
unity       & 12         & 22        & 14        & 19         & 7          & 8          \\
$^{233}$U   & -          & -         & 1         & -          & -          & -          \\
$^{235}$U   & 12         & -         & 28        & 19         & 13         & 4          \\
$^{239}$Pu  & -          & -         & -         & -          & 2          & -          \\
\hline
\end{tabular}
\end{center}
\end{table}

\subsection{Selection of datasets by energy and publication year}
The lower boundary of the incident energy for the present evaluation was unchanged from JENDL-4.0 (i.e., 10~keV for $^{233,235}$U and $^{239,241}$Pu, and 100~keV for $^{238}$U and  $^{240}$Pu)
while the upper boundary was extended from 20~MeV to 200~MeV.
All experimental datasets including data points in this energy range and published no earlier than 1970 ($^{233,238}$U and $^{239,240,241}$Pu) or 1980 ($^{235}$U) were extracted from the EXFOR Master File~\footnote{
A single ASCII file collecting all up-to-date EXFOR entries maintained by the IAEA Nuclear Data Section~\cite{Henriksson2008Art}
}.
We excluded the datasets digitized from article figure images, superseded by other datasets, and measured with neutrons from slowing-down time spectrometers and nuclear explosions.

\subsection{Exclusion of duplicated datasets}
A dataset published in an article may be revised and published again.
Usually the superseded dataset is flagged by {\sc spsdd} in the EXFOR entry,
and we can avoid double counting of the same measurement by excluding the flagged one.
However, identification of two datasets from the same measurement was not always trivial.
Two major examples are discussed below.

\subsubsection{Exclusion of duplicated datasets from Karlsruhe isochronous cyclotron}
There are several EXFOR entries providing the $^{235,238}$U and $^{239}$Pu(n,f) cross sections measured by Cierjacks et al. at the Karlsruhe isochronous cyclotron.
From this work,
the $^{235,238}$U absolute cross sections and their ratio were published separately~\cite{Leugers1976235U,Cierjacks1976Measurements} and compiled in EXFOR 20943 and 20409, respectively.
Their relation was discussed after the NEANDC/NEACRP Specialists Meeting on Fast Neutron Fission Cross Sections of U-233, U-235, U-238, and Pu-239 (28--30 June 1976, Argonne National Laboratory) between the editors of the meeting proceedings (W.P.~Poenitz and A.B.~Smith) and Cierjacks, and a note was added to Ref.~\cite{Leugers1976235U} to clarify that the ratio was derived from the absolute cross sections.
Similarly, the $^{235}$U and $^{239}$Pu absolute cross sections and their ratio from this work were published separately~\cite{Kari1978Measurement,Cierjacks1976Measurements} and compiled in EXFOR 20786 and 20409, respectively,
for which an author confirmed that each of the Pu isotopes was measured only once in this experimental work
and the 1976 and 1978 data should be based on the same raw data~\cite{Leugers2021}.

Finally, we adopted only the $^{238}$U/$^{235}$U cross section ratio in EXFOR 20409.002 and the $^{239}$Pu/$^{235}$U cross section ratio in EXFOR 20786.005 among several $^{238}$U and $^{239}$Pu datasets in these EXFOR entries,
and discarded their absolute cross sections to be free from the n-p scattering reference cross section used for normalization.

\subsubsection{Exclusion of duplicated data points from TUD-KRI collaboration}
Another major complication was selection of the final results from the absolute cross sections measured with the associated particle method by the Technische Universit\"{a}t Dresden (TUD) and Kholpin Radium Institute (KRI).
This question was introduced as an open problem in a review of the GMA database~\cite{Pronyaev2003Standards}.
The tables of Ref.~\cite{Dushin1983bStatistical} show there are several measurements for the same target nuclide and incident energy in some cases,
and this makes selection of the data points without double counting very difficult.
We traced the history of the measurements at TUD, KRI and ZFK (Zentralinstitut f\"{u}r Kernforschung, Rossendorf) by reviewing all journal articles and reports compiled in CINDA (Comprehensive Index of Nuclear Reaction Data)~\cite{OECD2007CINDA},
and selected the data points listed in Table~\ref{tab:TUDKRI} as inputs to our analysis.
See Ref.~\cite{Otuka2022Input} for more details.
\begin{table}[hbtp]
\caption{Absolute fission cross sections (in barns) from the TUD-KRI collaboration and adopted in present evaluation. ``$E_n$" gives the incident energy in~MeV and ``Lab." gives the location of the experiment.
The underline indicates presence of several cross sections at the same energy from the same laboratory.
The parenthesized value refers to the last digits of the uncertainty value (e.g., 1.240(24) means 1.240$\pm$0.024).
The incident energy value may slightly vary depending on the article when the same measurement is published twice or more.
}
\label{tab:TUDKRI}
\begin{center}
\begin{tabular}{rccccc}
\hline
    &      &\multicolumn{4}{c}{Target nuclide}\\
\cline{3-6}
$E_n$ &Lab.&$^{233}$U                             &$^{235}$U                                           &$^{238}$U                             &$^{239}$Pu                            \\
\hline                                                                                                                                                             
 2.6  &TUD &                                      &1.240(24)~\cite{Merla1991Absolute}                  &                                      &                                      \\
 4.5  &ROS &                                      &1.094(24)~\cite{Merla1991Absolute}                  &                                      &                                      \\
 4.8  &ROS &                                      &                                                    &0.562(17)~\cite{Merla1991Absolute}    &1.773(33)~\cite{Merla1991Absolute}    \\
 5.1  &ROS &                                      &                                                    &0.554(13)~\cite{Merla1991Absolute}    &                                      \\
 8.2  &ROS &                                      &                                                    &1.041(33)~\cite{Merla1991Absolute}    &                                      \\
 8.5  &ROS &                                      &\underline{1.855(44)~\cite{Merla1991Absolute}}      &                                      &2.395(40)~\cite{Merla1991Absolute}    \\
 8.5  &ROS &                                      &\underline{1.74(11)~\cite{Arlt1981aAbsoluteFission}}        &                                      &                                      \\
14.7  &TUD &2.244(42)~\cite{Dushin1983bStatistical}&2.096(24)~\cite{Merla1991Absolute}                  &1.194(22)~\cite{Alkhazov1979Fission}  &                                      \\
18.8  &ROS &                                      &2.068(50)~\cite{Merla1991Absolute}                  &1.363(43)~\cite{Merla1991Absolute}    &2.473(59)~\cite{Merla1991Absolute}    \\
\hline                                                                                                                               
 1.9  &KRI &1.93(7)~\cite{Shpakov1989Absolute}    &1.28(3)~\cite{Kalinin1991Correction}                &                                      &2.01(5)~\cite{Shpakov1989Absolute}    \\
 2.5  &KRI &                                      &1.27(3)~\cite{Kalinin1991Correction}                &                                      &                                      \\
14.0  &KRI &                                      &2.0840(362)~\cite{Dushin1983bStatistical}            &                                      &                                      \\
14.5  &KRI &                                      &2.1010(365)~\cite{Dushin1983bStatistical}            &                                      &                                      \\
14.7  &KRI &2.254(45)~\cite{Dushin1983bStatistical}&\underline{2.0960(289)~\cite{Dushin1983bStatistical}}&1.171(23)~\cite{Dushin1983bStatistical}&\underline{2.309(30)~\cite{Dushin1983bStatistical}}\\
14.7  &KRI &                                      &\underline{2.0755(361)~\cite{Dushin1983bStatistical}}&                                      &\underline{2.349(45)~\cite{Dushin1983bStatistical}}\\
14.7  &KRI &                                      &\underline{2.1348(308)~\cite{Dushin1983bStatistical}}&                                      &                                      \\
14.7  &KRI &                                      &\underline{2.0714(299)~\cite{Dushin1983bStatistical}}&                                      &                                      \\
\hline
\end{tabular}
\end{center}
\end{table}

\subsection{Modification to EXFOR entries}
\label{sec:unofficial}
Our guiding principle is to use the datasets in the official EXFOR entries without any further corrections.
To utilize additional information provided by the authors but cannot be in the official EXFOR entries,
we modified a few EXFOR entries for the present evaluation:
\begin{itemize}
\item $^{233,235,238}$U and $^{239}$Pu by Dushin et al.~\cite{Dushin1983bStatistical} (EXFOR 51001
\footnote{
EXFOR entry numbers starting from 5 are introduced to distinguish the modified EXFOR entries from the official EXFOR entries.
An ASCII file collecting these modified EXFOR entries is available as a Supplemental Material.
}
)\\
Prepared to collect the absolute fission cross sections measured by the associated particle method at KRI assuming that the KRI data points tabulated in Ref.~\cite{Dushin1983bStatistical} are the final ones (See also Table~\ref{tab:TUDKRI}).
The covariance matrices published in this reference were converted to the correlation coefficients, and used as inputs to our evaluation.
\item $^{238}$U/$^{235}$U by Poenitz et al.~\cite{Poenitz1972Measurements} (EXFOR 51002)\\
Prepared to merge three datasets EXFOR 10232.002, 003 and 004 to one dataset to include their correlation recorded by Poenitz in the GMA database entry \#816 and \#818.
\item $^{238}$U/$^{235}$U by Paradela et al.~\cite{Paradela2015High} (EXFOR 51005)\\
Prepared to merge two datasets EXFOR 23269.003 and 004 to one dataset to include their correlation due to use of the same sample.
\item $^{235}$U by Amaducci et al.~\cite{Amaducci2019Measurement} (EXFOR 51006)\\
Prepared to merge two datasets EXFOR 23453.002 and 003 to one dataset to include their correlation due to use of the same $^{235}$U fission counts .
\item $^{233}$U and $^{239,241}$Pu by Blons et al.~\cite{Blons1971MeasurementAnalysis,Blons1970MesureHaute,Blons1971MeasurementFission}(EXFOR 51007, 51008, 51009)\\
Prepared to average the high-resolution cross sections in EXFOR 20001.002, 20446.002 and 20484.002 to reduce the number of data points to 50 points per decade.
Note that the two data points at the highest two energies in EXFOR 51008.002 ($^{233}$U averaged dataset) were removed due to the unreasonable energy dependence seen in the original dataset (EXFOR 20446.002).
\item $^{235}$U by Merla et al.~\cite{Merla1991Absolute} (EXFOR 51010)\\
Prepared to utilize the information on the statistical uncertainties in a preliminary report~\cite{Alkhazov1988New} together with the data points compiled from the final report (EXFOR 22304.006).
\end{itemize}

\subsection{Construction of experimental covariances}
\label{sec:const}
A critical step of the current evaluation is determination of the covariances of each experimental dataset.
We used several approaches depending on the available information in the EXFOR entry and source article.
A summary on the range and source of each partial uncertainty included in construction of the covariances is published elsewhere~\cite{Otuka2022Input}.
\begin{itemize}
\item
Correlation coefficients in EXFOR\\
Correlation coefficients are explicitly given under the keyword {\sc covariance} in the EXFOR entry for a few cases (EXFOR 
13169.002~\cite{Meadows1989Search},
14498.002~\cite{Casperson2018Measurement},
22211.002~\cite{Iwasaki1990Measurement},
22282.003.1 and 006.1~\cite{Manabe1988Measurements},
41112.002~\cite{Kalinin1991Correction},
51001.002-005~\cite{Dushin1983bStatistical},
51006.002~\cite{Amaducci2019Measurement}
)
and they were directly taken as inputs to our fitting.
\item
Correlated and uncorrelated partial uncertainties in EXFOR\\
When the correlated and uncorrelated partial uncertainties are in the EXFOR entry separately, 
we used them for estimation of the correlation coefficients.
The quadrature sum of the correlated and uncorrelated partial uncertainties exceeds the total uncertainty in EXFOR 23458.006~\cite{Stamatopoulos2020Investigation}.
For this dataset,
we discarded the uncorrelated partial uncertainty in the EXFOR entry,
treated the total uncertainty in the EXFOR entry as uncorrelated,
and combined it with the correlated partial uncertainty in the EXFOR entry.
\item
Correlated or uncorrelated partial uncertainty and total uncertainty in EXFOR\\
When the correlated or uncorrelated partial uncertainty is missing in the EXFOR entry but the total uncertainty is coded under the heading {\sc err-t} in the EXFOR entry, 
we derived the missing partial uncertainty by subtracting the other partial uncertainty from the total uncertainty assuming the quadrature sum rule.
The quadrature sum of the correlated or uncorrelated partial uncertainties exceeds the total uncertainty in some data points of seven datasets 
(EXFOR
10653.004~\cite{Behrens1977Measurements},
21764.004~\cite{Budtz-Jorgensen1981Neutron},
22321.006~\cite{Eismont1996Relative},
22698.005~\cite{Tovesson2004233Pa},
23078.002-003~\cite{Nolte2007Cross},
40506.002~\cite{Fursov1977aMeasurement}),
and these data points were excluded from the evaluation.
There is one more dataset creating the same problem in many points (EXFOR 22014.003~\cite{Kanda1986Measurement}),
and we treated the uncertainty declared as the total uncertainty in the EXFOR entry as uncorrelated.
The total uncertainty is absent for the data points below 10 keV in EXFOR 10267.002~\cite{Gwin1976Measurement}, and these data points were discarded.
\item
Correlated partial uncertainty and an energy-dependent uncertainty in EXFOR\\
In order to increase the number of usable datasets,
we treated the energy-dependent uncertainty not clearly declared as the total uncertainty (coded under the heading {\sc data-err} in EXFOR) as total in 10 datasets (EXFOR
20567.003-004~\cite{Szabo1970New},
20569.004~\cite{Szabo1971235U},
20570.003-004~\cite{Szabo1973Mesure},
30548.002-003~\cite{Khan1980New},
40673.004-005~\cite{Aleksandrov1983Neutron},
51010.002~\cite{Merla1991Absolute}
).
The uncorrelated partial uncertainty was obtained by subtraction of the correlated partial uncertainty from the energy-dependent uncertainty assuming the quadrature sum rule.
\item
Energy-dependent uncertainty in EXFOR with normalization uncertainty estimated by us\\
Sometimes a fully correlated (overall normalization) partial uncertainty can be easily inferred from the article text even if it cannot be in the official EXFOR entries due to the EXFOR policy.
When such information exists with an energy-dependent uncertainty in the EXFOR entry,
we estimated the normalization uncertainty and stored it in an input file rather than modification of the EXFOR entry.
The uncorrelated partial uncertainty was obtained by subtraction of the normalization uncertainty from the energy-dependent uncertainty in the EXFOR entry assuming the quadrature sum rule.
Table~\ref{tab:normunc} summarizes such normalization uncertainties estimated by us with their justifications.
\begin{table}[hbtp]
\caption{Normalization uncertainties not coded in the EXFOR entries but used in the current evaluation. ``Unc." gives the uncertainty in \%.}
\label{tab:normunc}
\begin{minipage}{\textwidth}
\begin{center}
\begin{tabular}{lllp{10cm}}
\hline
Target               &EXFOR \#     &Unc.   & Source of uncertainty\\
\hline                            
$^{233}$U            & 13890.004   & 1.0   & Uncertainty due to normalization to the thermal value~\cite{Wagemans1988Subthermal} (0.25\% from thermal normalization +0.7\% from point-wise uncertainty)\footnotemark
\\
$^{233}$U            & 51008.002   & 1.4   & Uncertainty in the thermal reference value~\cite{James1970Cross}\footnotemark\\
$^{238}$U            & 13169.003.2 & 1.0   & Normalization uncertainty (minimum 1\%, see 13169.002)\\
$^{240}$Pu           & 40673.004   & 0.15  & Uncertainty in the target half-life (6357+/-10 yr)\\
$^{241}$Pu           & 40673.005   & 1.4   & Uncertainty in the target half-life (14.4+/-0.2 yr)\\
$^{241}$Pu           & 51009.002   & 1.5   & Uncertainty in the thermal reference value~\cite{James1970Cross}\footnotemark\\
$^{233}$U/$^{235}$U  & 41432.003   & 0.74  &
Uncertainty in the fissile nucleus number ratio
(taken from a preliminary report of the same experiment~\cite{Shpak1980Measurement})\\
$^{239}$Pu/$^{235}$U & 20569.004   & 1.0   & Uncertainty in the sample mass
(0.6\% from $^{239}$Pu+0.8\% from $^{235}$U)\\
\hline 
\end{tabular}
\end{center}
\stepcounter{footnote}\footnotetext[1]{
Guber et al.~\cite{Guber2000New} adopt the 8.1-to-17.6~eV cross section integral of 965.2~eV-b determined by Wagemans et al.~(Table 1 of Ref.~\cite{Wagemans1988Subthermal}),
which is from their point-wise cross sections normalized to the thermal cross section of 531.14~b$\pm$0.25\%.
The uncertainty in the cross section integral originated from the energy dependence of the point-wise cross section is not mentioned by Wagemans et al.
We estimated it to 0.7\% by propagating Wagemans' point-wise uncertainty in EXFOR 22080.002 to the cross section integral.
}
\stepcounter{footnote}\footnotetext[2]{
Blons et al.~\cite{Blons1973High} adopt the 8.32-to-101.2~eV resonance integral of 168.31~b evaluated by James~\cite{James1970Cross}.
The uncertainty in the resonance integral (1.4\%) was estimated by us by propagating the uncertainty in each resonance integral in Table I of Ref.~\cite{James1970Cross}.
}
\stepcounter{footnote}\footnotetext[3]{
Blons et al.~\cite{Blons1973High} adopt the 20-to-70~eV cross section integral of 2367.5~eV-b evaluated by James~\cite{James1970Cross}.
The uncertainty in the cross section integral (1.5\%) was estimated by us by propagating the uncertainty in each group-wise cross section in Table IV of Ref.~\cite{James1970Cross}.
}
\end{minipage}
\end{table}
\item Special cases\\
The uncertainties due to statistics, detector efficiency and position are combined into one uncertainty in EXFOR 20618.003~\cite{Szabo1976Measurement}, and it was treated as uncorrelated.
The energy-dependent uncertainty without source specification in EXFOR 40483.002~\cite{Vorotnikov1976Sub} was assumed as uncorrelated,
and the dataset was treated as a shape dataset without correlated partial uncertainties.
\item Other cases\\
All datasets not belonging to the above-mentioned categories were excluded from the present evaluation.
\end{itemize}
In general, the lower boundary  was adopted as a constant uncertainty when only a range is known for the partial uncertainty.

It is not trivial to estimate the correlation coefficient when a partial uncertainty is neither uncorrelated (e.g., statistical uncertainty) nor fully correlated (e.g., normalization uncertainty).
Following ``Occam's Razor" strategy~\cite{Smith2012Experimental},
we treated all partial uncertainties other than those due to counting statistics (coded under the heading {\sc err-s} or with the correlation flag {\sc u} in EXFOR) as fully correlated uncertainties,
namely our estimation gives the upper limit of the actual correlation coefficient.
Presence of correlation between two datasets from the same experimental work was found in a few EXFOR entries~\cite{Poenitz1972Measurements, Paradela2015High,Amaducci2019Measurement}, and we merged such datasets to a single dataset to take into account the known correlation appropriately (see Sect.\ref{sec:unofficial}).
Any two datasets in our experimental database were therefore treated as independent each other.
\section{Least-squares fitting}
The least-squares fitting was performed by SOK, which updated the prior estimates of the cross sections (taken from evaluated data libraries) by including the experimental dataset one-by-one from our experimental database
\footnote{
A video (animated GIF file) recording the data update in chronological order of experimental works is available as a Supplemental Material.
}.

\subsection{Formalism}
\begin{figure}[hbtp]
\centering\includegraphics[clip,angle=-90,width=0.7\linewidth]{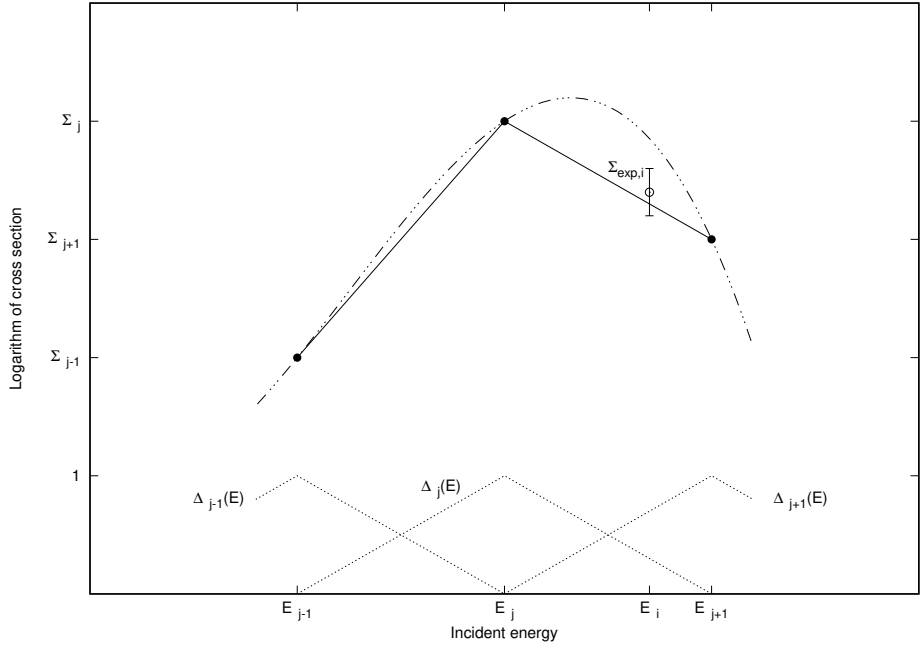}
\caption{Fitting of the logarithm of experimental cross section $\Sigma_{\rm exp}$ to Schmittroth's roof function $\Delta$~\cite{Schmittroth1980Finite}.}
\label{fig:roof}
\end{figure}

The SOK code adopts Schmittroth's roof function~\cite{Schmittroth1980Finite} (Fig.~\ref{fig:roof}) as the model to express the logarithm of the cross section $\Sigma(E)=\ln\sigma(E)$ by introducing $n$ energy nodes between $E_1$ and $E_n$:
\begin{equation}
\Sigma(E)=\sum_{j=1}^n \Sigma_j \Delta_j(E)
\end{equation}
with
\begin{equation}
\Delta_j(E)=
\left\{
\renewcommand{\arraystretch}{1.2}
\begin{array}{ll}
\dfrac{E-E_{j-1}}{E_j-E_{j-1}}& (E_{j-1} \le E \ \le E_j)\\
\dfrac{E_{j+1}-E}{E_{j+1}-E_j}& (E_j \le E \ < E_{j+1})\\
0                             & \textrm{otherwise}       \\
\end{array}
\renewcommand{\arraystretch}{1}
\right.
,
\end{equation}
which is equivalent to fitting to the first-order B-spline function~\cite{Uenohara1983Simultaneous,Uenohara1984Reviews}.
The fitting parameter $\Sigma_j=\ln \sigma_j\,(j=1,n)$ is the logarithm of the evaluated cross section at $E_j$.
The logarithm of an experimental cross section at $E_i$ ($E_j \le E_i < E_{j+1}$) is related with the fitting parameters by
\begin{equation}
\Sigma_{{\rm exp},i}=
\Sigma_j \Delta_j(E_i) + \Sigma_{j+1} \Delta_{j+1}(E_i) + \delta_i=
\Sigma_j c_{i,j}       + \Sigma_{j+1} c_{i,j+1}         + \delta_i
\end{equation}
with the residual of fitting $\delta_i$ and 
\begin{equation}
\left\{
\begin{array}{ll}
c_{i,j}  &=(E_{j+1}-E_i)/(E_{j+1}-E_j) \\
c_{i,j+1}&=(E_i    -E_j)/(E_{j+1}-E_j) \\
\end{array}
\right.
.
\end{equation}
See Ref.~\cite{Kawano2000Simultaneous,Kawano2000Evaluation} for further details about SOK such as the least-squares solution and treatment of the cross section ratio.
The evaluated cross sections in the JENDL-4.0 library (below 20~MeV) and JENDL-4.0/HE library~\cite{Kunieda2016Overview} (above 20~MeV) were adopted as the prior estimates of the parameters except for the $^{233}$U fission cross section above 20~MeV, which is not in the JENDL-4.0/HE library and the evaluation by Yabshits et al.~\cite{Yavshits2001Multiconfiguration} was adopted instead.
Note that the high energy part of the fission cross sections in the JENDL-4.0/HE library is from the JENDL/HE-2007 library~\cite{Watanabe2011Status}.
The uncertainties in all prior parameters were set to 50\% without correlation among them.

In the present work, the original version of SOK used for JENDL-3.3 and JENDL-4.0 evaluations was slightly extended to treat shape datasets (i.e., datasets in arbitrary unit).
The logarithm of an experimental cross section in arbitrary unit $\Sigma'_{{\rm exp},i}$ may be related with the fitting parameters by
\begin{equation}
\Sigma'_{{\rm exp},i}=\Sigma_j c_{i,j} + \Sigma_{j+1} c_{i,j+1} -\ln a +\delta_i ,
\end{equation}
where $a$ is an additional fitting parameter normalizing the cross section in the arbitrary unit to the absolute one and we took $1.0\pm0.5$ as its prior estimate.
In addition to the dataset EXFOR 40483.002 (see Sect.~\ref{sec:const}),
we also treated the $^{238}$U dataset in EXFOR 14529.002~\cite{Miller2015Measurement} as a shape dataset excluding the normalization uncertainty (4.1\%) since the dataset is originally normalized to IAEA Neutron Data Standards 2006 (IAEA-2006)~\cite{IAEA2007International,Carlson2009International} at 130~MeV,
and we would like to maintain our evaluation free from this normalization.
The parameter $a$ and its uncertainty were estimated in the present least-squares analysis for normalization within the fission cross sections of the six nuclides.
Such shape datasets are also included in the GMA database in the IAEA evaluation where normalization is done taking into account not only fission standards but also more fundamental standards of light nuclides,
and one could expect more reasonable estimate of the normalization parameters than those estimated in the present evaluation.

It is obvious from the formalism that the evaluated cross sections at the lower and upper boundaries of the energy range for evaluation must be determined with the experimental data points not only inside but also outside of the energy range.
Namely, we need extra energy nodes below the lower boundary and above the upper boundary~\cite{Schmittroth1980Finite}.
In order to take into account this effect, 
we added a few extra nodes to include the data points of each dataset above 7~keV ($^{233,235}$U, $^{239,241}$Pu) or 70~keV (~$^{238}$U, $^{240}$Pu) and below 250~MeV.

\subsection{Revision of experimental database after preliminary fitting (LANSCE data)}
\begin{figure}[hbtp]
\centering\includegraphics[clip,angle=-90,width=0.7\linewidth]{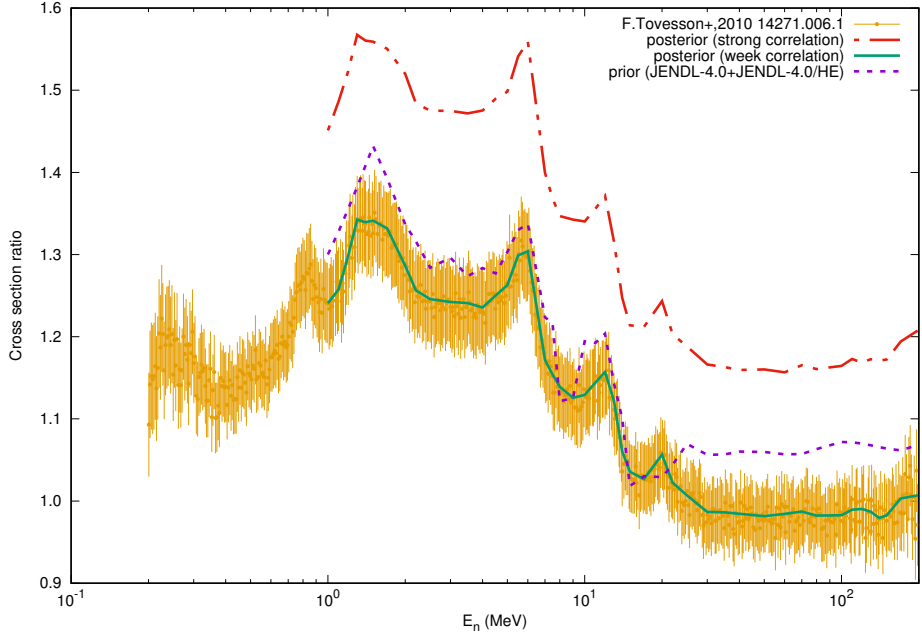}
\caption{Fitting to the LANSCE $^{241}$Pu(n,f)/$^{235}$U(n,f) cross section ratio (EXFOR 14271.006.1)~\cite{Tovesson2010Cross} with strong correlation (i.e., all partial uncertainties other than the statistical uncertainty are treated as fully correlated) and weak correlation (i.e., all partial uncertainties other than the normalization uncertainty are treated as uncorrelated).
Note that the experimental data points below 1~MeV were discarded in the present evaluation.}
\label{fig:tovesson}
\end{figure}

\begin{figure}[hbtp]
\centering\includegraphics[clip,angle=-90,width=0.7\linewidth]{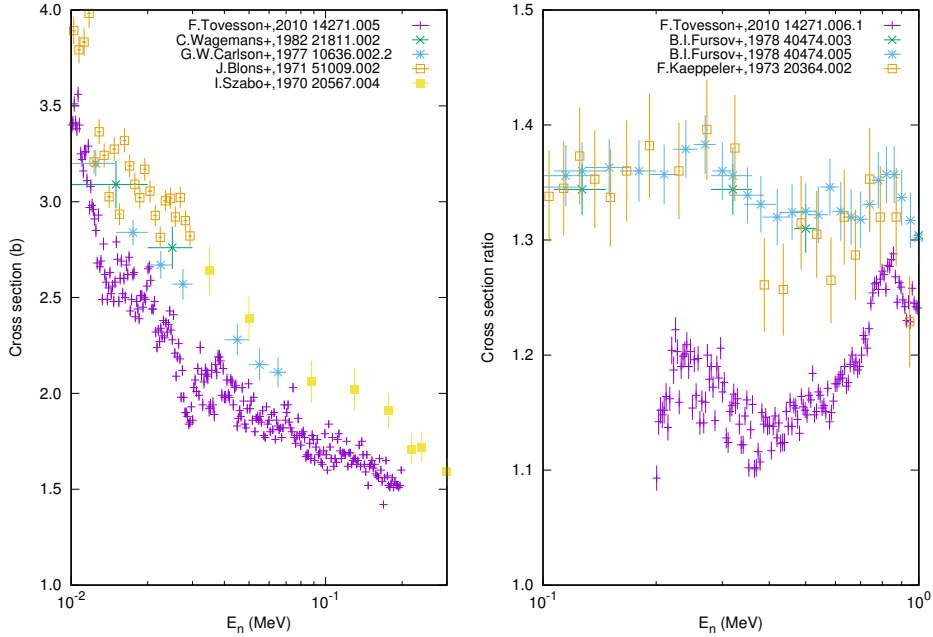}
\caption{Comparison of the LANSCE $^{241}$Pu(n,f) cross section (EXFOR 14271.005) and $^{241}$Pu(n,f)/$^{235}$U(n,f) cross section ratio (EXFOR 14271.006.1)~\cite{Tovesson2010Cross} with other datasets.
The cross section ratio from LANSCE is normalized to the ratio at 25.3~meV by Tovesson et al.
The absolute cross section is converted from the ratio by using the ENDF/B-VII library by Tovesson et al.}
\label{fig:tovesson1MeV}
\end{figure}

Tovesson et al. published fission cross sections measured at LANSCE~\cite{Tovesson2009Neutron,Tovesson2010Cross,Tovesson2014Fast} for various target nuclides after JENDL-4.0 evaluation.
They are very useful for the present evaluation since these measurements cover the neutron energy up to 200~MeV and compiled in the EXFOR library with the cross section ratios and energy-dependent partial uncertainties in general.
Following their instruction~\cite{Tovesson2015Uncertainty},
we initially constructed the correlation coefficients of these datasets assuming that the energy-dependent and independent ``systematic" uncertainties are fully correlated.
This is consistent with our general procedure but it sometimes leads to very strong correlation coefficients and strange results as demonstrated in Fig.~\ref{fig:tovesson}.
The result becomes more reasonable if we weaken the correlation by treating all partial uncertainties other than the normalization uncertainty as uncorrelated,
and we adopted this procedure for all LANSCE datasets measured by Tovesson et al. (EXFOR 14271.003.1 and 006.1~\cite{Tovesson2010Cross}, 14402.003 and 009~\cite{Tovesson2014Fast}).
Note that we excluded their (1) $^{241}$Pu/$^{235}$U dataset below 1~MeV (EXFOR 14271.006.1~\cite{Tovesson2010Cross}) and $^{241}$Pu dataset (EXFOR 14271.005~\cite{Tovesson2010Cross}) because they do not agree with almost all data points of the other datasets within the error bars between 20~keV and 800~keV as shown in Fig.~\ref{fig:tovesson1MeV}, and (2) $^{239,240}$Pu datasets (EXFOR 14223.002~\cite{Tovesson2009Neutron} and 14271.002~\cite{Tovesson2010Cross}) because their ratios to $^{235}$U are not available from the authors.
A theoretical study~\cite{Neudecker2021Informing} concludes that the most of the discrepant $^{241}$Pu datasets between 0.1 and 2~MeV can be covered by changing the fission barrier height of $^{242}$Pu by only 150~keV and one cannot exclude a discrepant dataset by such theoretical consideration.

\clearpage
\section{Results}
\begin{figure}[hbtp]
\centering\includegraphics[clip,angle=-90,width=0.8\linewidth]{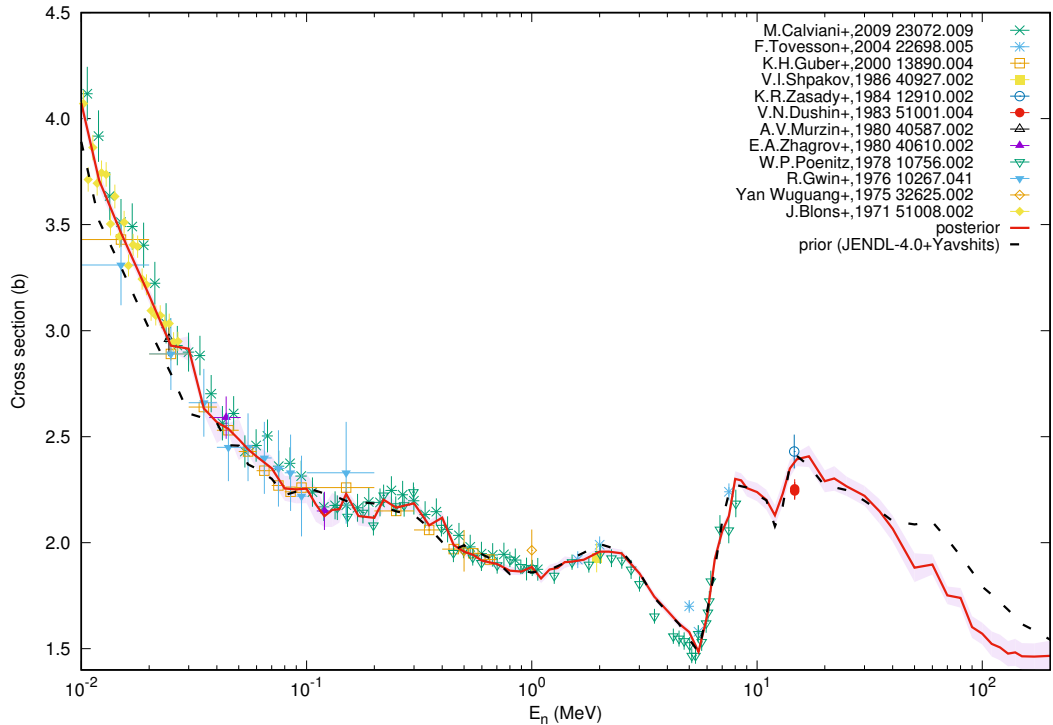}
\caption{$^{233}$U fission prior and posterior cross sections with the experimental cross sections used for evaluation\cite{
Calviani2009High,
Tovesson2004233Pa,
Guber2000New,
Shpakov1989Absolute,
Zasadny1984Measurement,
Dushin1983bStatistical,
Murzin1980Measurement,
Zhagrov1980Fission,
Poenitz1978Absolute,
Gwin1976Measurement,
Yan1975Measurement,
Blons1971MeasurementAnalysis}.
The prior cross section is taken from JENDL-4.0 (below 20 MeV) and Yavshits' evaluation (above 20 MeV).
}
\label{fig:U233log}
\end{figure}

\begin{figure}[hbtp]
\centering\includegraphics[clip,angle=-90,width=0.8\linewidth]{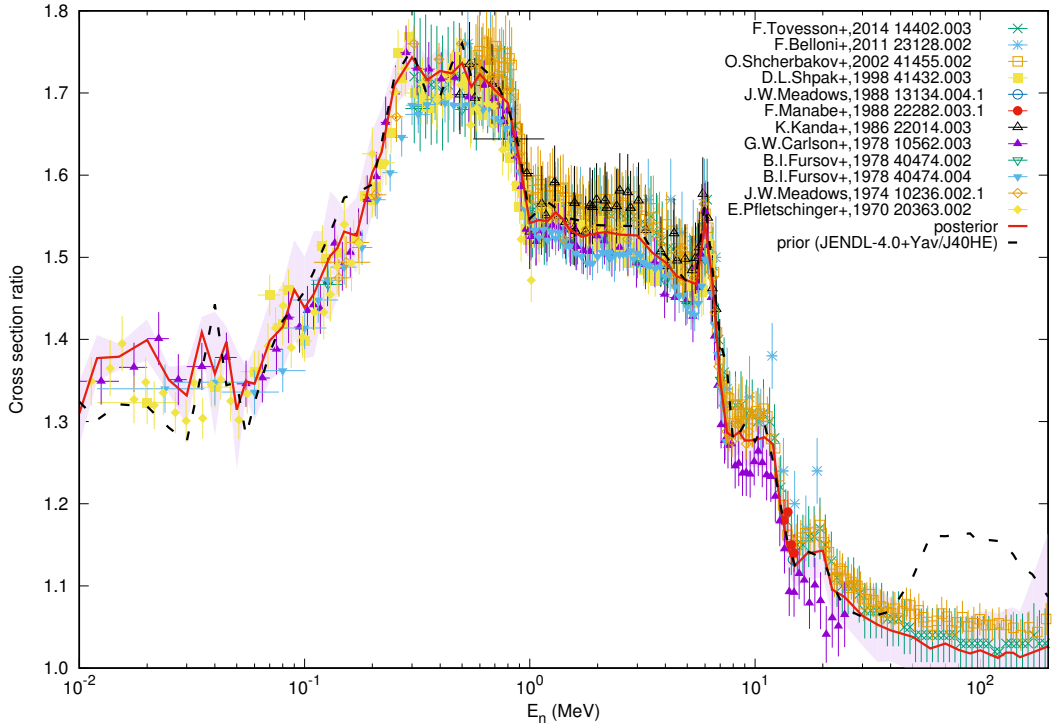}
\caption{$^{233}$U/$^{235}$U fission prior and posterior cross section ratios with the experimental cross section ratios used for evaluation\cite{
Tovesson2014Fast,
Belloni2011Neutron,
Shcherbakov2002Neutron,
Shpak1998Angular,
Meadows1988Fission,
Manabe1988Measurements,
Kanda1986Measurement,
Carlson1978aMeasurement,
Fursov1978Measurement,
Meadows1974Ratio,
Pfletschinger1970Measurement}.
The prior cross section is taken from JENDL-4.0 (below 20 MeV), Yavshits' evaluation (above 20 MeV, $^{233}$U) or JENDL-4.0/HE (above 20 MeV, $^{235}$U).
}
\label{fig:U233U235log}
\end{figure}

\begin{figure}[hbtp]
\centering\includegraphics[clip,angle=-90,width=0.8\linewidth]{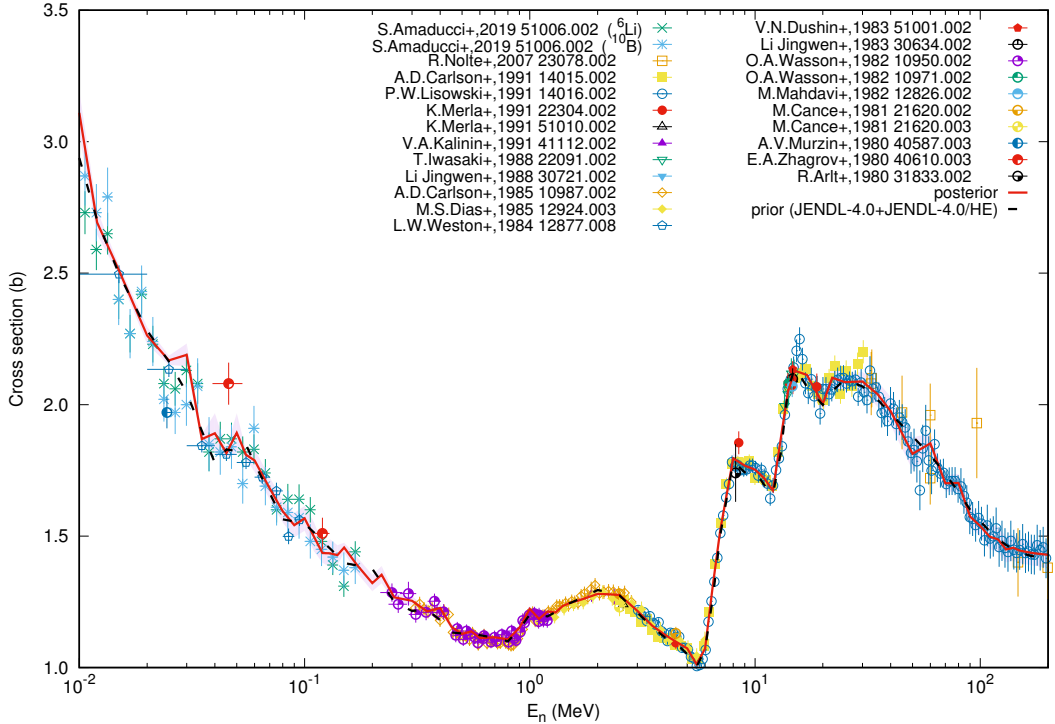}
\caption{$^{235}$U fission prior and posterior cross sections with the experimental cross sections used for evaluation\cite{
Amaducci2019Measurement,
Nolte2007Cross,
Carlson1992Measurements,
Lisowski1991Fission,
Merla1991Absolute,
Kalinin1991Correction,
Iwasaki1988Measurement,
Li1988Absolute,
Carlson1985Absolute,
Dias1985Application,
Weston1984Subthreshold,
Dushin1983bStatistical,
Li1983Absolute,
Wasson1982Absolute,
Wasson1982Measurement,
Mahdavi1983Measurements,
Cance1981Measures,
Murzin1980Measurement,
Zhagrov1980Fission,
Arlt1981aAbsoluteFission}.
The prior cross section is taken from JENDL-4.0 (below 20 MeV) and JENDL-4.0/HE (above 20 MeV).
In the legends of Amaducci et al.'s datasets, $^6$Li and $^{10}$B denote the cross sections normalized with the $^6$Li(n,t)$^4$He and $^{10}$B(n,$\alpha$)$^7$Li standard cross sections, respectively. }
\label{fig:U235log}
\end{figure}

\begin{figure}[hbtp]
\centering\includegraphics[clip,angle=-90,width=0.8\linewidth]{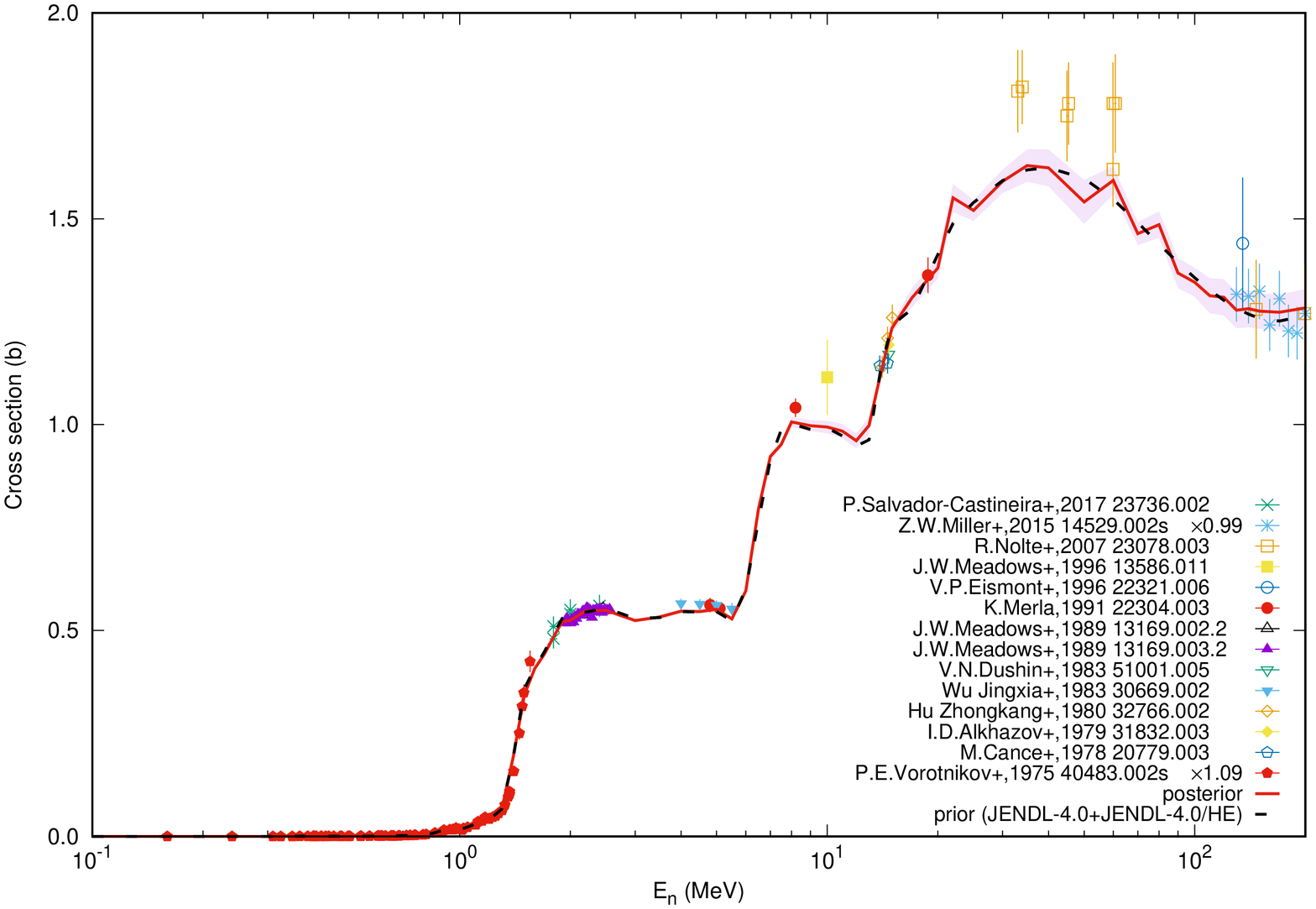}
\caption{$^{238}$U fission prior and posterior cross sections with the experimental cross sections used for evaluation\cite{
Salvador-Castineira2017Absolute,
Miller2015Measurement,
Nolte2007Cross,
Meadows1996Measurement,
Eismont1996Relative,
Merla1991Absolute,
Meadows1989Search,
Dushin1983bStatistical,
Wu1983Measurement,
Hu1980Measurement,
Alkhazov1979Fission,
Cance1978Absolute,
Vorotnikov1976Sub}.
The prior cross section is taken from JENDL-4.0 (below 20 MeV) and JENDL-4.0/HE (above 20 MeV).
}
\label{fig:U238log}
\end{figure}

\begin{figure}[hbtp]
\centering\includegraphics[clip,angle=-90,width=0.8\linewidth]{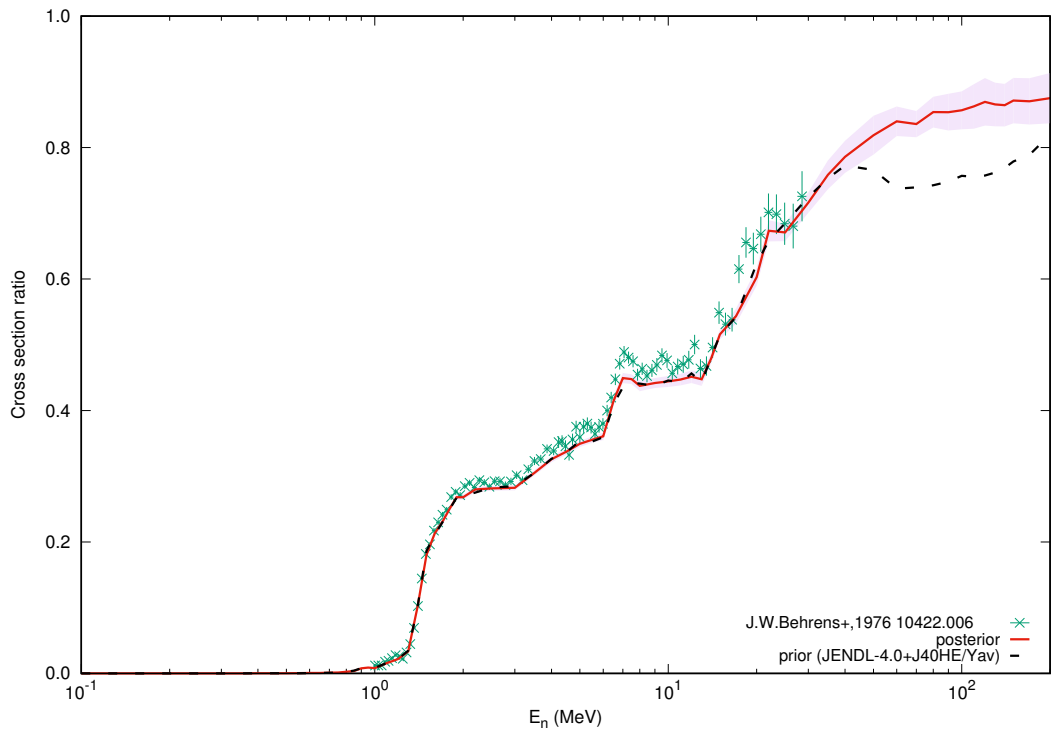}
\caption{$^{238}$U/$^{233}$U fission prior and posterior cross section ratios with the experimental cross section ratios used for evaluation\cite{
Behrens1976Measurements}.
The prior cross section is taken from JENDL-4.0 (below 20 MeV), JENDL-4.0/HE (above 20 MeV, $^{238}$U) or Yavshits' evaluation (above 20 MeV, $^{233}$U).
}
\label{fig:U238U233log}
\end{figure}

\begin{figure}[hbtp]
\centering\includegraphics[clip,angle=-90,width=0.8\linewidth]{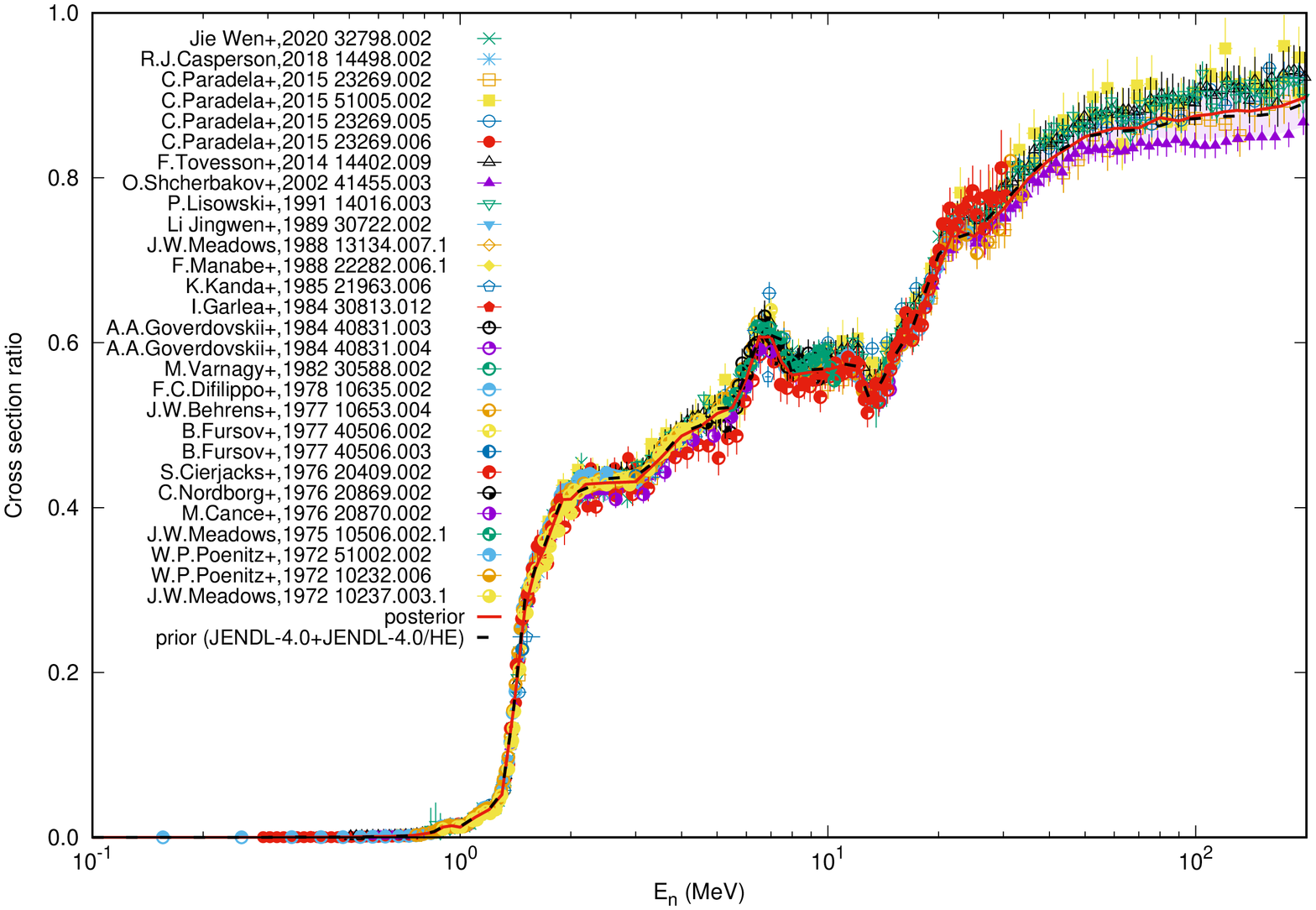}
\caption{$^{238}$U/$^{235}$U fission prior and posterior cross section ratios with the experimental cross section ratios used for evaluation\cite{
Wen2020Measurement,
Casperson2018Measurement,
Paradela2015High,
Tovesson2014Fast,
Shcherbakov2002Neutron,
Lisowski1991Fission,
Li1989Ratio,
Meadows1988Fission,
Manabe1988Measurements,
Kanda1985Measurement,
Garlea1984Measuring,
Goverdovskii1984aMeasurement,
Varnagy1982New,
Difilippo1978Measurement,
Behrens1977Measurements,
Fursov1977aMeasurement,
Cierjacks1976Measurements,
Nordborg1976Fission,
Cance1976Measurements,
Meadows1975Ratio,
Poenitz1972Measurements,
Meadows1972Ratio}.
The prior cross section is taken from JENDL-4.0 (below 20 MeV) and JENDL-4.0/HE (above 20 MeV).
}
\label{fig:U238U235log}
\end{figure}

\begin{figure}[hbtp]
\centering\includegraphics[clip,angle=-90,width=0.8\linewidth]{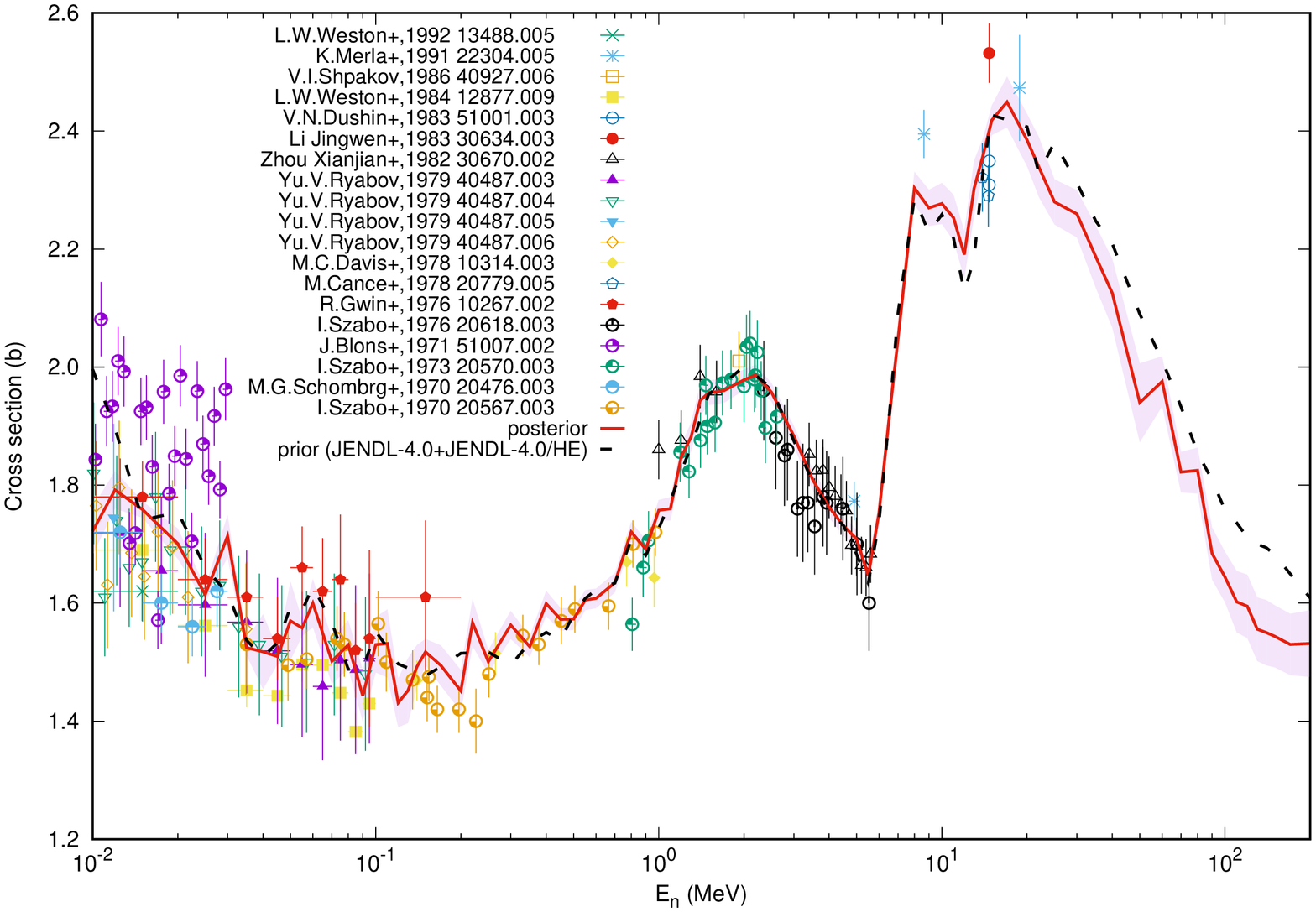}
\caption{$^{239}$Pu fission prior and posterior cross sections with the experimental cross sections used for evaluation\cite{
Weston1992High,
Merla1991Absolute,
Shpakov1989Absolute,
Weston1984Subthreshold,
Dushin1983bStatistical,
Li1983Absolute,
Zhou1983Fast,
Ryabov1979Measurements,
Davis1978Absolute,
Cance1978Absolute,
Gwin1976Measurement,
Szabo1976Measurement,
Blons1970MesureHaute,
Szabo1973Mesure,
Schomberg1970Ratio,
Szabo1970New}.
The prior cross section is taken from JENDL-4.0 (below 20 MeV) and JENDL-4.0/HE (above 20 MeV).
}
\label{fig:Pu239log}
\end{figure}

\begin{figure}[hbtp]
\centering\includegraphics[clip,angle=-90,width=0.8\linewidth]{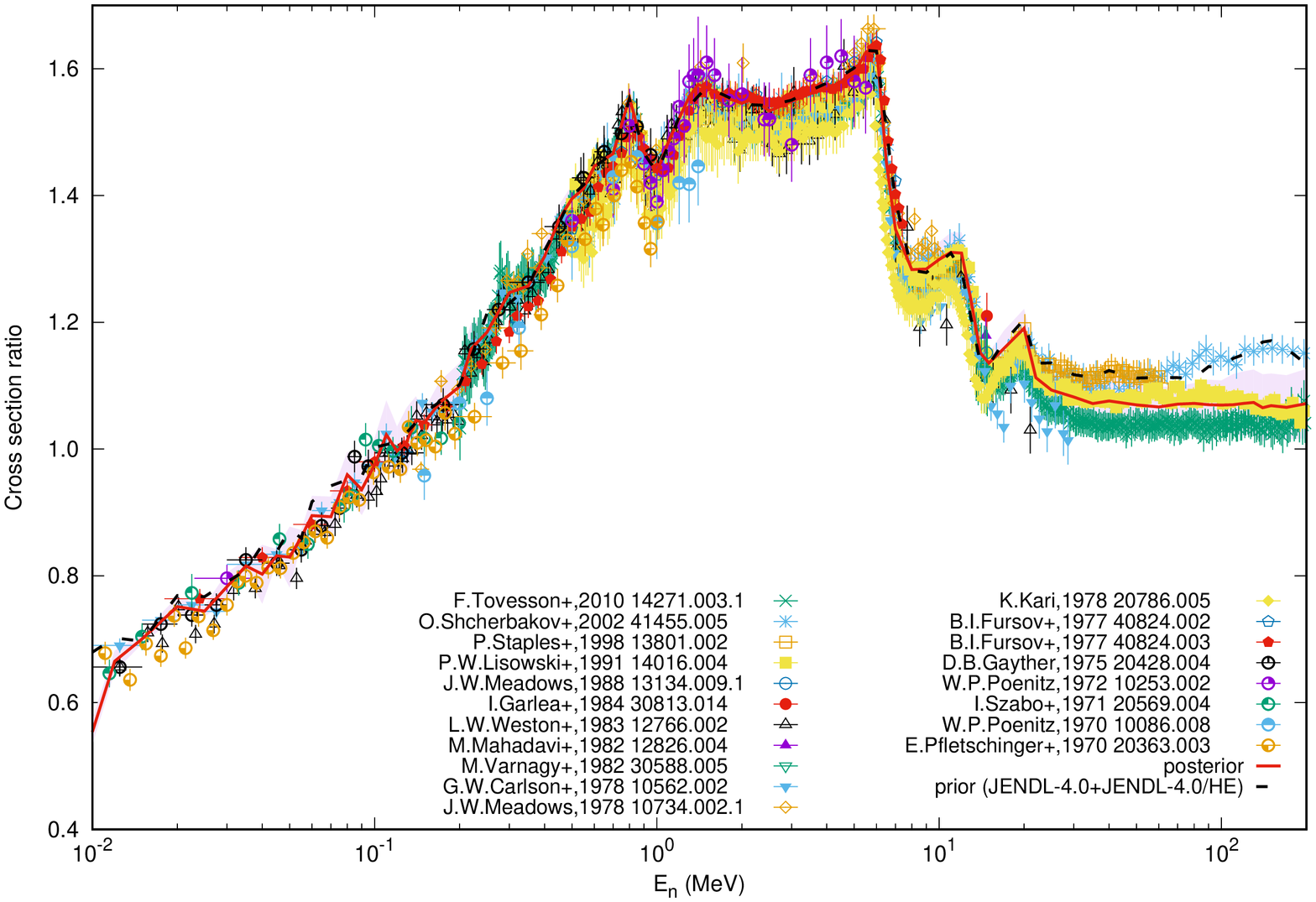}
\caption{$^{239}$Pu/$^{235}$U fission prior and posterior cross section ratios with the experimental cross section ratios used for evaluation\cite{
Tovesson2010Cross,
Shcherbakov2002Neutron,
Staples1998Neutron,
Lisowski1991Fission,
Meadows1988Fission,
Garlea1984Measuring,
Weston1983Neutron,
Mahdavi1983Measurements,
Varnagy1982New,
Carlson1978aMeasurement,
Meadows1978Fission,
Kari1978Measurement,
Fursov1977bMeasurement,
Gayther1975Measurement,
Poenitz1972Additional,
Szabo1971235U,
Poenitz1970Measurement,
Pfletschinger1970Measurement}.
The prior cross section is taken from JENDL-4.0 (below 20 MeV) and JENDL-4.0/HE (above 20 MeV).
}
\label{fig:Pu239U235log}
\end{figure}

\begin{figure}[hbtp]
\centering\includegraphics[clip,angle=-90,width=0.8\linewidth]{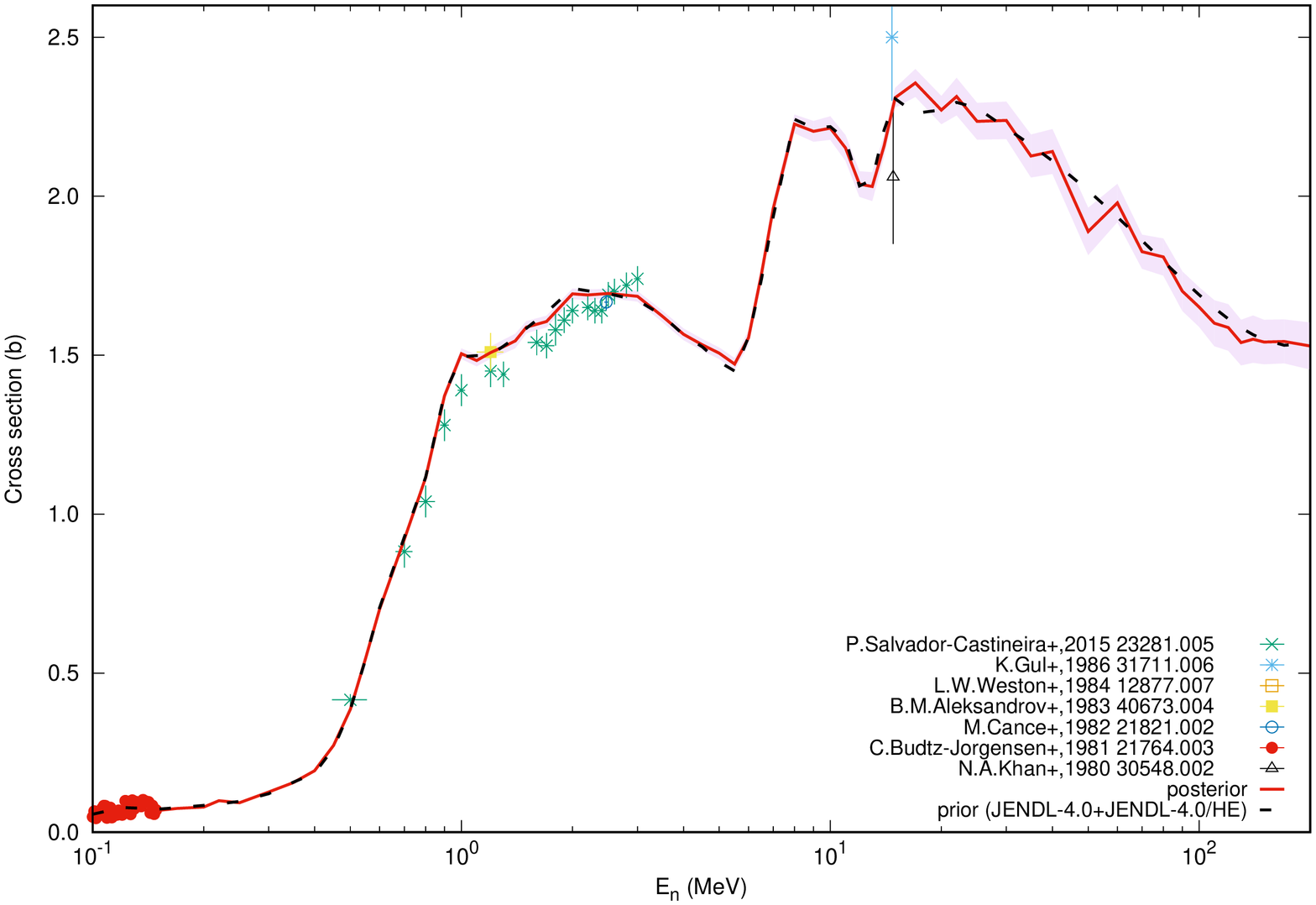}
\caption{$^{240}$Pu fission prior and posterior cross sections with the experimental cross sections used for evaluation\cite{
Salvador-Castineira2015Neutron,
Gul1986Measurements,
Weston1984Subthreshold,
Aleksandrov1983Neutron,
Cance1983Mesures,
Budtz-Jorgensen1981Neutron,
Khan1980New}.
The prior cross section is taken from JENDL-4.0 (below 20 MeV) and JENDL-4.0/HE (above 20 MeV).
}
\label{fig:Pu240log}
\end{figure}

\begin{figure}[hbtp]
\centering\includegraphics[clip,angle=-90,width=0.8\linewidth]{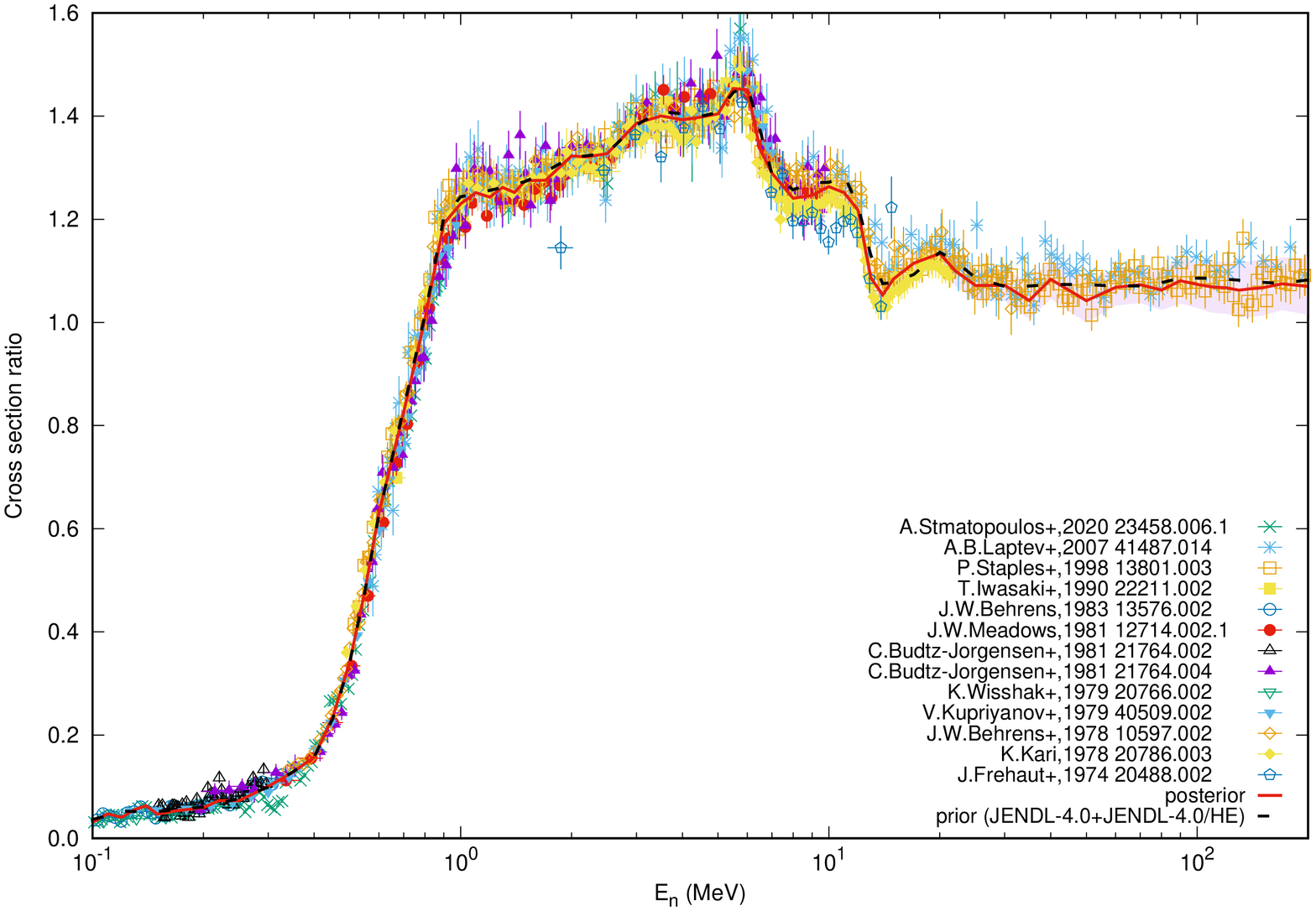}
\caption{$^{240}$Pu/$^{235}$U fission prior and posterior cross section ratios with the experimental cross section ratios used for evaluation\cite{
Stamatopoulos2020Investigation,
Laptev1998Fast,
Staples1998Neutron,
Iwasaki1990Measurement,
Behrens1983Measurement,
Meadows1981Fission,
Budtz-Jorgensen1981Neutron,
Wisshak1979Measurement,
Kupriyanov1979Measurement,
Behrens1978Measurements,
Kari1978Measurement,
Frehaut1974Mesures}.
The prior cross section is taken from JENDL-4.0 (below 20 MeV) and JENDL-4.0/HE (above 20 MeV).
}
\label{fig:Pu240U235log}
\end{figure}

\begin{figure}[hbtp]
\centering\includegraphics[clip,angle=-90,width=0.8\linewidth]{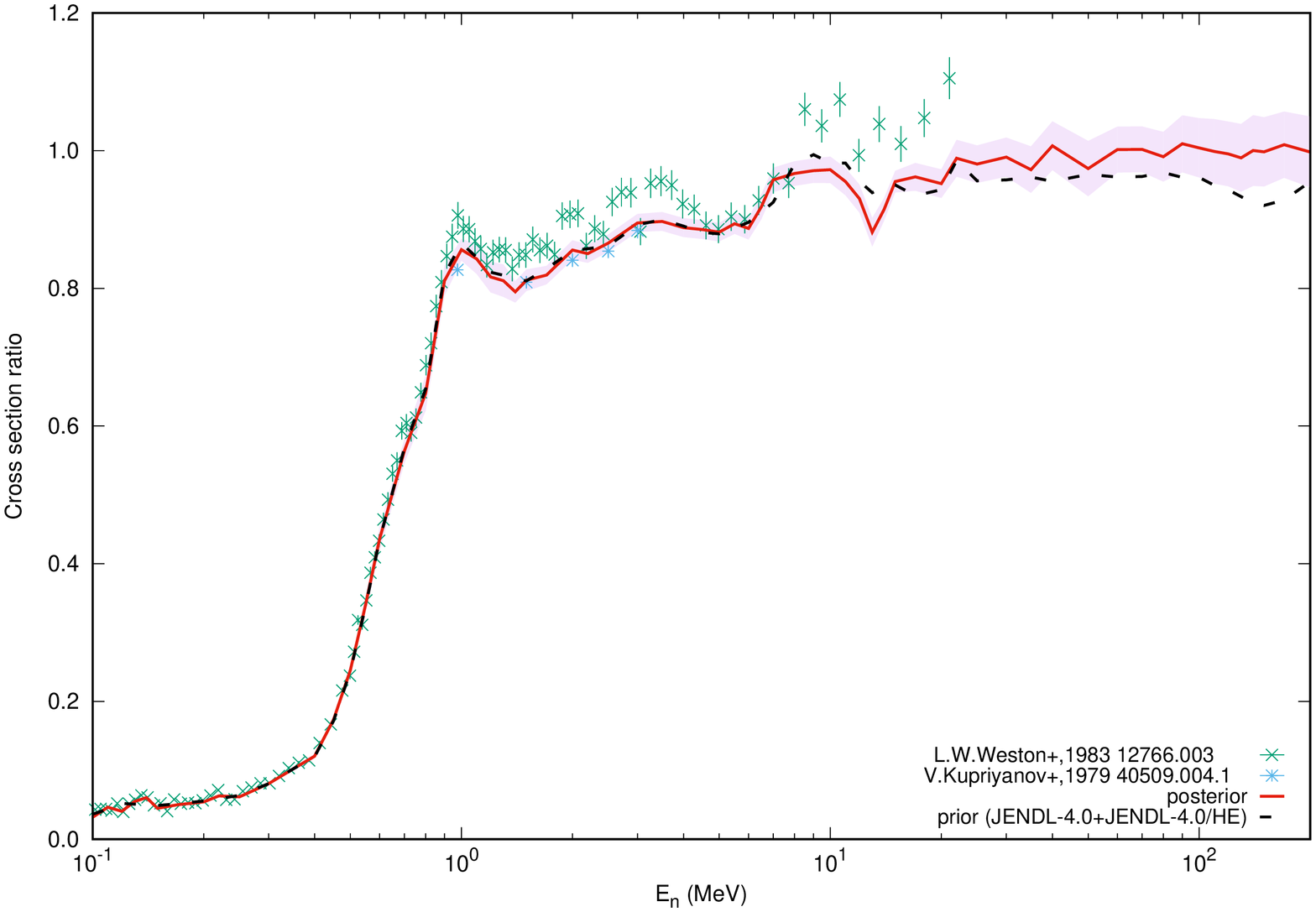}
\caption{$^{240}$Pu/$^{239}$Pu fission prior and posterior cross section ratios with the experimental cross section ratios used for evaluation\cite{
Weston1983Neutron,
Kupriyanov1979Measurement}.
The prior cross section is taken from JENDL-4.0 (below 20 MeV) and JENDL-4.0/HE (above 20 MeV).
}
\label{fig:Pu240Pu239log}
\end{figure}

\begin{figure}[hbtp]
\centering\includegraphics[clip,angle=-90,width=0.8\linewidth]{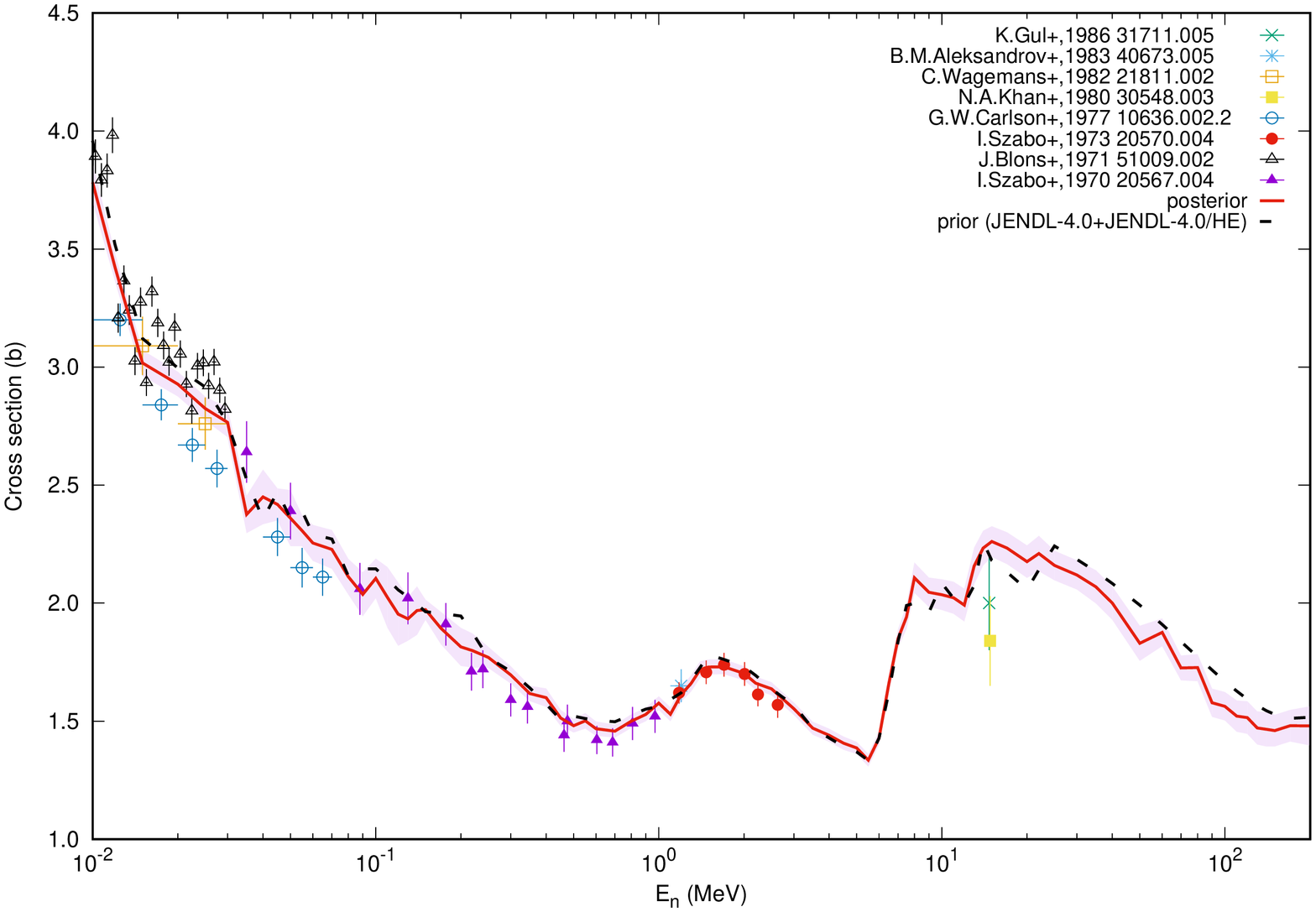}
\caption{$^{241}$Pu fission prior and posterior cross sections with the experimental cross sections used for evaluation\cite{
Gul1986Measurements,
Aleksandrov1983Neutron,
Wagemans1982241Pu,
Khan1980New,
Carlson1977Measurement,
Szabo1973Mesure,
Blons1970MesureAnalyse,
Blons1971MeasurementFission,
Szabo1970New}.
The prior cross section is taken from JENDL-4.0 (below 20 MeV) and JENDL-4.0/HE (above 20 MeV).
}
\label{fig:Pu241log}
\end{figure}

\begin{figure}[hbtp]
\centering\includegraphics[clip,angle=-90,width=0.8\linewidth]{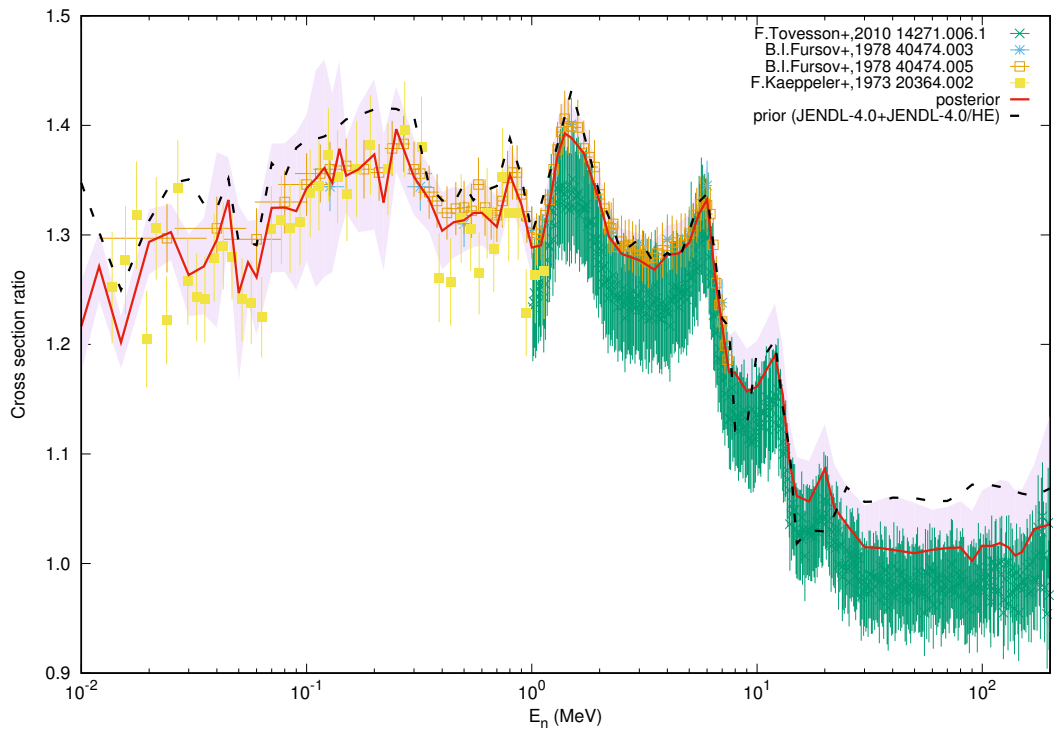}
\caption{$^{241}$Pu/$^{235}$U fission prior and posterior cross section ratios with the experimental cross section ratios used for evaluation\cite{
Tovesson2010Cross,
Fursov1978Measurement,
Kaeppeler1973Measurement}.
The prior cross section is taken from JENDL-4.0 (below 20 MeV) and JENDL-4.0/HE (above 20 MeV).
}
\label{fig:Pu241U235log}
\end{figure}

The number of the experimental data points used for the present evaluation was 7379, and the number of the fitting parameters was 497 including two fitting parameters for shape data normalization (EXFOR 14529.002 and 40483.002).
With these parameters, the reduced chi-square (chi-square divided by the degree-of-freedom) is 4.00 if we compare the experimental database with the evaluated cross sections in the whole energy range for fitting (7 or 70~keV to 250~MeV).
The reduced chi-square slightly increases to 4.45 if we lower the upper energy boundary from 250~MeV to 20~MeV, which is still smaller than the reduced chi-square calculated with the same experimental database and the prior cross sections (JENDL-4.0) in the same energy range (6.57).
The newly evaluated (posterior) cross sections from final fitting are plotted with the prior cross sections and experimental cross sections in Figs.~\ref{fig:U233log} to \ref{fig:Pu241U235log}
\footnote{
The evaluated cross sections in an ASCII file are available as a Supplemental Material.
}.
The band accompanying the newly evaluated cross section in each figure shows the external uncertainty in the evaluated cross section modelled by the roof function.
See Sect.\ref{sec:unc} for more details.

\clearpage
\subsection{Comparison with JENDL-4.0 and IAEA Neutron Data Standards 2017}
Here we discuss comparison of the newly evaluated cross sections with the cross sections in the JENDL-4.0 library and IAEA Neutron Data Standards 2017 (IAEA-2017).
The JENDL-4.0 cross sections were evaluated by the SOK code but with another experimental database, which was constructed from EXFOR by a different approach and also does not include a number of new experimental datasets.
The IAEA-2017 cross sections were evaluated for $^{235,238}$U and $^{239}$Pu by the GMA code as described in Sect.~\ref{sec:majorlibs}.

\begin{figure}[hbtp]
\centering\includegraphics[clip,angle=-90,width=0.8\linewidth]{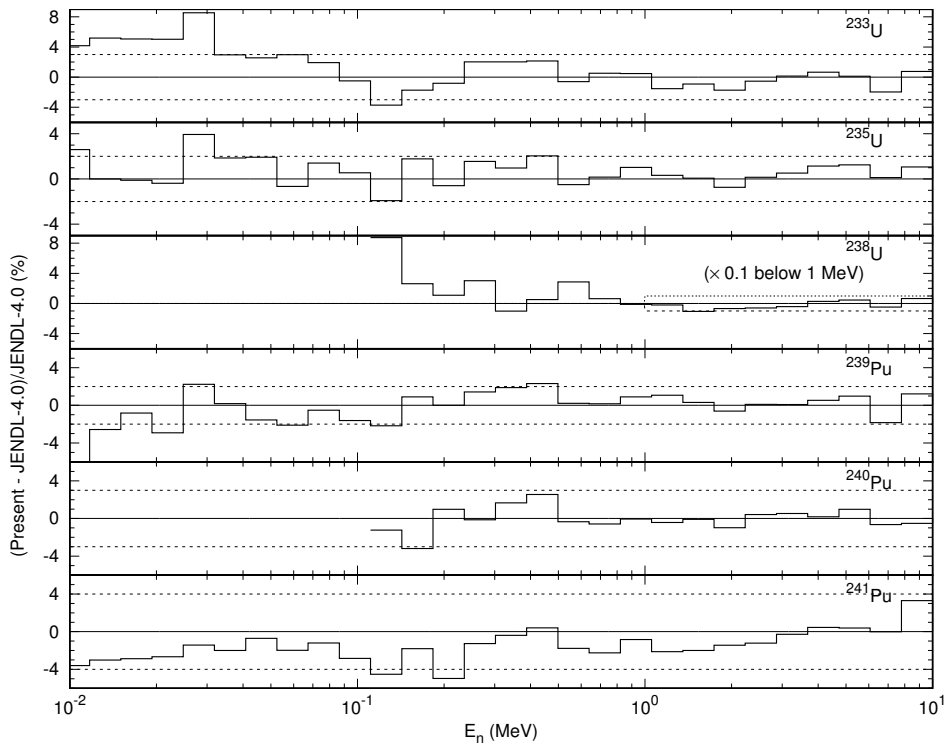}
\caption{Difference of the newly evaluated cross sections from the JENDL-4.0 cross sections in the JFS-3 70-group structure.}
\label{fig:diffJ40}
\end{figure}

Figure~\ref{fig:diffJ40} shows difference of the newly evaluated cross sections from the JENDL-4.0 cross sections in the 70-group structure defined in the JFS-3 (JAERI-Fast Set Ver.3) format~\cite{Takano1978JAERI,Takano1989Revision}.
The group-wise cross sections were calculated by replacing the fission cross sections in the JENDL-4.0 library with the newly evaluated ones in the ENDF-6 format by DeCE~\cite{Kawano2019Dece} and processing by PREPRO2019~\cite{Cullen2019PREPRO}.
Roughly speaking, the change from JENDL-4.0 is within 4\% for $^{241}$Pu, 3\% for $^{233}$U and $^{240}$Pu, and 2\% for $^{235}$U and $^{239}$Pu.
For $^{238}$U, the change is small (within 1\%) above 1~MeV but large in the sub-threshold region.

We observe collective increase of the cross sections of $^{233}$U, $^{235}$U and $^{239}$Pu in the 24th group (24.8~keV to 31.8~keV).
Amaducci et al.~\cite{Amaducci2019Measurement} report that JENDL-4.0 underestimates the $^{235}$U cross section from their measurement about 4 and 5\% in 30-60 and 60-100~keV region, respectively.
Figure~\ref{fig:diffJ40} shows that the newly evaluated $^{235}$U cross section is also increased by 2--4\% from the JENDL-4.0 cross section in the 22th to 24th group (24.8 to 52.5~keV) and less increase in the 19th to 20th group (67.4--111~keV).

\begin{figure}[hbtp]
\centering\includegraphics[clip,angle=-90,width=0.8\linewidth]{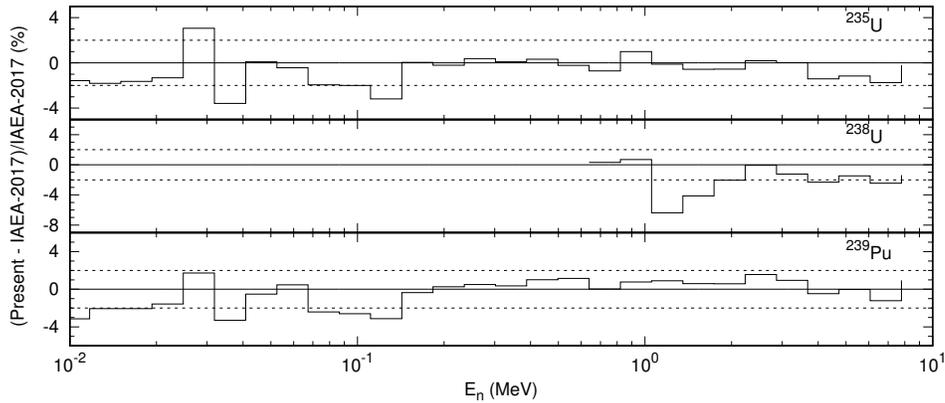}
\caption{Difference of the newly evaluated cross sections from the IAEA Neutron Data Standards 2017 cross sections in the JFS-3 70-group structure.}
\label{fig:diffI17}
\end{figure}

Figure~\ref{fig:diffI17} shows difference of the newly evaluated cross sections from the IAEA-2017 cross sections in the JFS-3 70-group structure.
They agree within 2\% for all three nuclides with some exceptions.
The collective increase in the $^{235}$U and $^{239}$Pu cross sections in the 24th group (24.8--31.8~keV) seen in comparison with the JENDL-4.0 library appears again in this comparison.
We performed a trial fit excluding some datasets which seem relevant to this deviation such as $^{233}$U by Calviani et al. (23072.009~\cite{Calviani2009High}), $^{235}$U by Amaducci et al. (51006.002~\cite{Amaducci2019Measurement}) and $^{239}$Pu by Blons et al (51007.002~\cite{Blons1970MesureHaute},
but it did not eliminate the cross section structures seen in this energy region.
Collective decrease in the $^{235}$U and $^{239}$Pu cross sections in the 18th group (111--143~keV) is also commonly seen in the comparison with the JENDL-4.0 and IAEA-2017.
The difference in the $^{238}$U cross section in the threshold region (1--2~MeV) is also remarkable,
and this could be due to the small bump at $\sim$1.2~MeV visible in some high-resolution experimental datasets (e.g., EXFOR 23269.007 measured by Paradela et al.~\cite{Paradela2015High}).

\subsection{Validation by californium-252 spontaneous fission neutron spectrum averaged cross sections}
\label{sec:cf252}
As a validation of the newly evaluated cross sections in the prompt fission neutron energy region,
we derived the californium-252 (Cf-252) spontaneous fission neutron spectrum averaged cross sections (SACS) from the newly evaluated cross sections,
and compared them with those derived from the cross sections in the other evaluated data libraries and experimental ones.
We updated the JENDL-4.0 files by replacing the fission cross sections with the newly evaluated ones in the ENDF-6 format by using DeCE,
and converted the cross sections in the updated files to the SAND-II 725 energy group structure by PREPRO2019, which were then averaged over the Cf-252 spontaneous fission neutron spectrum evaluated by Mannhart~\cite{Mannhart1987Evaluation,Mannhart1989Status} and compiled in the IRDFF-II library in the same group structure.

\begin{figure}[hbtp]
\centering\includegraphics[clip,angle=-90,width=0.8\linewidth]{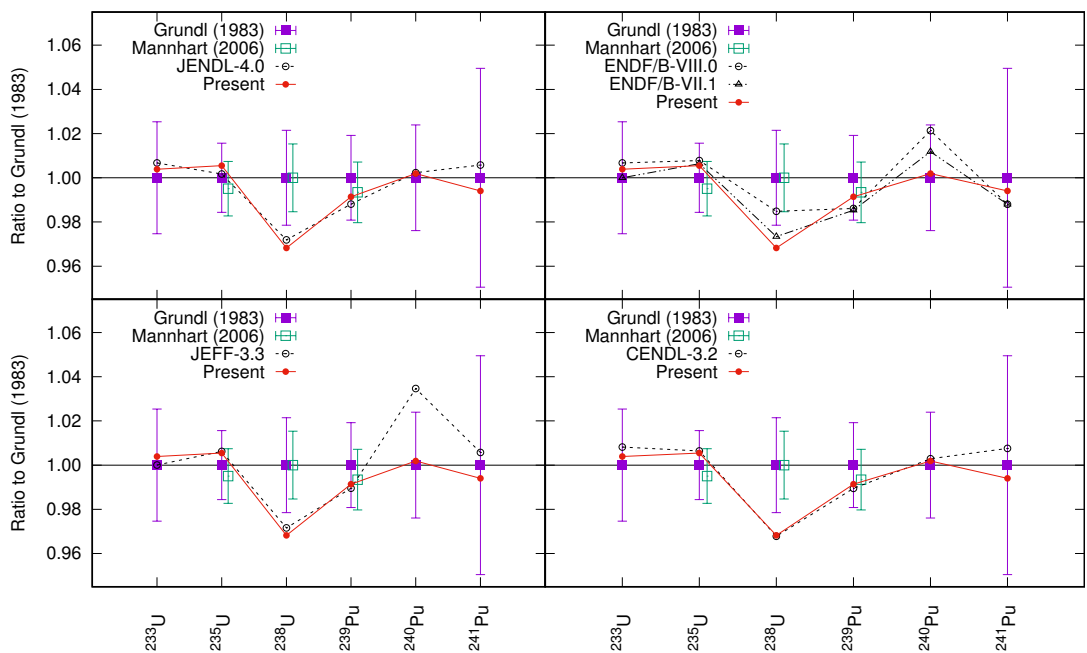}
\caption{Californium-252 spontaneous fission neutron spectrum averaged cross sections relative to those measured by Grundl and Gilliam~\cite{Grundl1983Fission} for the present evaluation, JENDL-4.0, ENDF/B-VIII.0, ENDF/B-VII.1, JEFF-3.3 and CENDL-3.2 evaluations as well as Mannhart's recommendation~\cite{Mannhart2006Response}.
Note that ENDF/B-VIII.0 adopts JENDL-4.0 for $^{233}$U, and JEFF-3.3 adopts JENDL-4.0 for $^{241}$Pu.}
\label{fig:sacs-cf252}
\end{figure}

Figure~\ref{fig:sacs-cf252} shows the ratios of the evaluated SACS to the SACS measured by Grundl and Gilliam~\cite{Grundl1983Fission} for the present evaluation as well as the JENDL-4.0, ENDF/B-VIII.0, ENDF/B-VII.1~\cite{Chadwick2011ENDF}, JEFF-3.3 and CENDL-3.2 evaluations.
In addition to the SACS derived from these evaluations,
the SACS recommended by Mannhart~\cite{Mannhart2006Response} are also shown for $^{233}$U, $^{235}$U and $^{239}$Pu.
The SACS from the present evaluation are also summarized in Table~\ref{tab:sacs-cf252} along with those measured by Grundl et al. and recommended by Mannhart.
This figure shows that a similar degree of agreement with the measured and recommended SACS is achieved by the present and JENDL-4.0 evaluation.
The SACS from the present evaluation agrees with the measured and recommended SACS within error bars except for $^{238}$U.
We tried various fitting (e.g., by changing the energy grid structure in the threshold region for $^{238}$U),
but it was not possible to improve their agreement.
The underestimation of the measured and recommended SACS for $^{238}$U is seen not only in JENDL evaluations but also in many other evaluations.
Among the ENDF/B evaluations, ENDF/B-VII.1 SACS is close to the present SACS while the ENDF/B-VIII.0 SACS is within the error bar of the SACS measured by Grundl and Gilliam, and also agrees with the lower boundary of Mannhart's recommended SACS.

\begin{figure}[hbtp]
\centering\includegraphics[clip,angle=-90,width=0.8\linewidth]{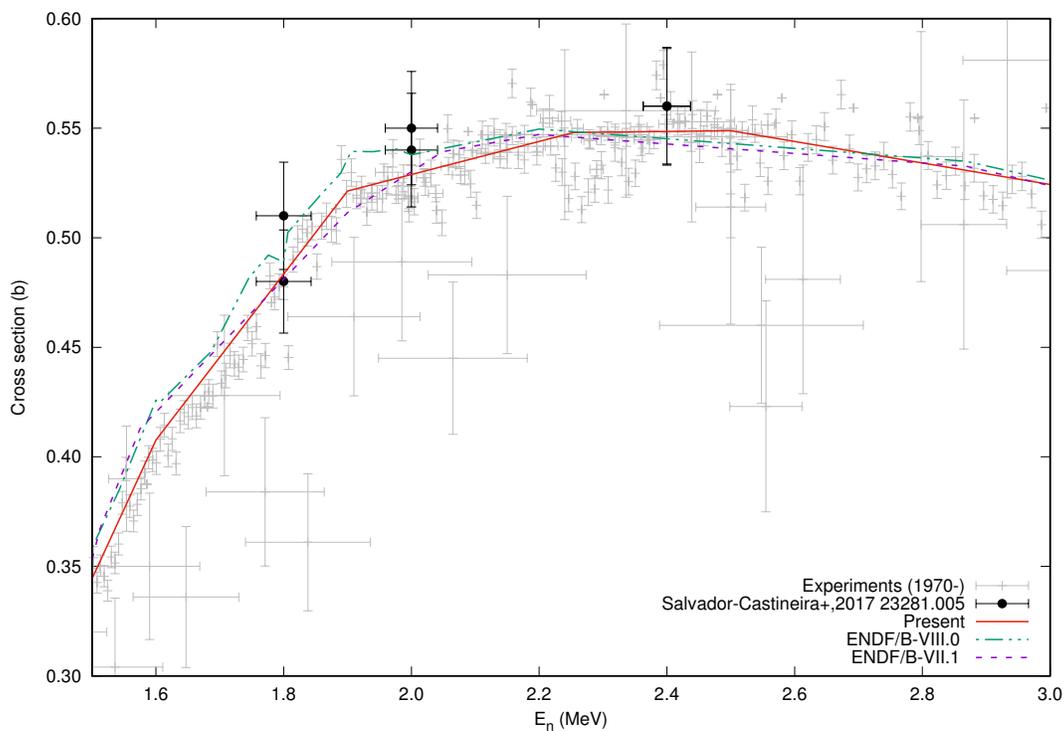}
\caption{
Comparison of the $^{238}$U(n,f) cross sections from the present evaluation with the evaluated cross sections in ENDF/B-VIII.0 and ENDF/B-VII.1 as well as absolute cross sections in the EXFOR library compiled from the articles published no earlier than 1970.
The solid circles show the data points from a recent (2017) absolute measurement at NPL~\cite{Salvador-Castineira2017Absolute}.
%
}
\label{fig:U238exfenf}
\end{figure}

Since the Cf-252 SACS has large sensitivity with respect to the energy dependent cross sections around 2~MeV,
we compared the present $^{238}$U point-wise cross section with those in ENDF/B-VII.1 and ENDF/B-VIII.0 in Fig.~\ref{fig:U238exfenf},
where all absolute experimental cross sections compiled in the EXFOR library published no earlier than 1970 are also plotted.
This figure shows the ENDF/B-VIII.0 gives much higher cross section than the ENDF/B-VII.1 and present evaluation between 1.7 and 2.0~MeV.
In this energy region, the ENDF/B-VIII.0 cross section is higher than the majority of the experimental cross sections, 
but more consistent with the recently published cross sections from an absolute measurement at National Physical Laboratory (NPL) with neutron flux determined by a long counter~\cite{Salvador-Castineira2017Absolute}.
The ENDF/B-VIII.0 cross section is based on IAEA-2017 which difference from the present evaluation in this energy region is visible in Fig.~\ref{fig:diffI17}.

\begin{table}[hbtp]
\caption{Californium-252 spontaneous fission neutron spectrum averaged cross sections (mb).}
\label{tab:sacs-cf252}
\begin{center}
\begin{tabular}{lllllll}
\hline
                                     & $^{233}$U  & $^{235}$U & $^{238}$U      & $^{239}$Pu & $^{240}$Pu & $^{241}$Pu \\
\hline
Present                              & 1900       & 1223       & 316           & 1808       & 1340       & 1606       \\
Grundl~\cite{Grundl1983Fission}      & 1893$\pm$48& 1216$\pm$19& 326$ \pm$6.5  & 1824$\pm$35& 1337$\pm$32& 1616$\pm$80\\
Mannhart~\cite{Mannhart2006Response} &            & 1210$\pm$15& 325.7$\pm$5.3 & 1812$\pm$25&            &            \\
\hline
\end{tabular}
\end{center}
\end{table}

\subsection{Validation by $\Sigma\Sigma$ spectrum averaged cross sections}
The neutron field of a coupled thermal/fast uranium and boron carbide spherical assembly $\Sigma\Sigma$-ITN (Bucharest) was developed for benchmark of the fast neutron dosimetry~\cite{Garlea1981SigmaSigma-ITN}.
Spectrum averaged cross sections of various dosimetry reactions were measured at this facility~\cite{Garlea1978Measuring,Garlea1980Integral,Garlea1981SigmaSigma-ITN},
and have been utilized for validation of the dosimetry cross section libraries such as IRDF-2002~\cite{Shibata2002Average} and IRDFF-II~\cite{Trkov2019IRDFF-II}.
The $\Sigma\Sigma$ neutron spectrum (average energy of 730~keV~\cite{Trkov2020IRDFF-II}) is softer than the Cf-252 spontaneous fission neutron spectrum,
and we calculated the newly evaluated cross sections averaged over the $\Sigma\Sigma$ spectrum for validation complementary to the Cf-252 SACS validation.
The $\Sigma\Sigma$ neutron spectrum tabulated by Fabry~\cite{Fabry1975Secondary} in a 135 energy group structure converted to the SAND-II 725 energy group structure~\cite{Simakov2021} was used for tabulation of the spectrum averaged cross sections for the IRDFF-II full summary report~\cite{Trkov2019IRDFF-II},
and we also used the same 725 energy group spectrum for our validation.
The evaluated cross sections updated from JENDL-4.0 in the 725 energy group structure were prepared by DeCE and PREPRO2019.

\begin{figure}[hbtp]
\centering\includegraphics[clip,angle=-90,width=0.8\linewidth]{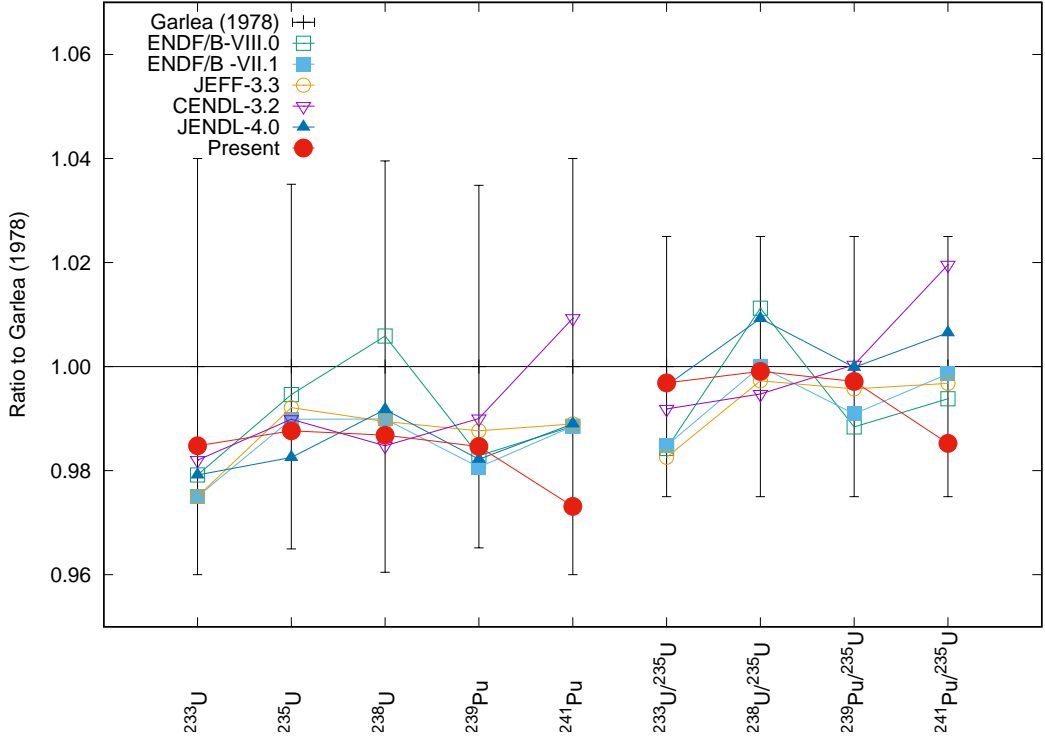}
\caption{
Ratios to the $\Sigma\Sigma$ neutron spectrum averaged cross sections measured by G\^{a}rlea et al.~\cite{Garlea1978Measuring} for those derived from the cross sections of our present evaluation as well as JENDL-4.0, ENDF/B-VIII.0, ENDF/B-VII.1, JEFF-3.3 and CENDL-3.2 evaluations.
Note that ENDF/B-VIII.0 adopts JENDL-4.0 for $^{233}$U and JEFF-3.3 adopts JENDL-4.0 for $^{241}$Pu.}
\label{fig:sacs-sigsig}
\end{figure}

Figure~\ref{fig:sacs-sigsig} shows the ratios of the evaluated SACS to the SACS measured at $\Sigma\Sigma$-ITN by G\^{a}rlea et al.~\cite{Garlea1981SigmaSigma-ITN} for the present evaluation as well as the JENDL-4.0, ENDF/B-VIII.0, ENDF/B-VII.1, JEFF-3.3 and CENDL-3.2 evaluations.
The SACS from the present evaluation are also summarized in Table~\ref{tab:sacs-sigsig} along with those measured by G\^{a}rlea et al.
The figure shows that the spectrum averaged cross sections of all evaluations agree with the measured ones within the experimental error bar,
which is originated from statistics (about 0.5\%), sample mass (1.6\% to 3.0\%), run-to-run monitor level, and corrections.
G\v{a}rlea et al. 
use their cross section ratios to the $^{235}$U cross section (1512$\pm$53~mb) for comparison with other $\Sigma\Sigma$ measurements.
Similarly, we also added the ratios from the present evaluation and libraries to Fig.~\ref{fig:sacs-sigsig} for comparison with their experimental ratios.
It shows the ratios of the cross sections to the $^{235}$U cross section from the present evaluation also agree with the measured ones.
%

\begin{table}[hbtp]
\caption{$\Sigma\Sigma$ neutron spectrum averaged cross sections (mb).}
\label{tab:sacs-sigsig}
\begin{center}
\begin{tabular}{llllll}
\hline
                                     & $^{233}$U  & $^{235}$U  & $^{238}$U     & $^{239}$Pu & $^{241}$Pu \\
\hline
Present                              & 2290       & 1493       & 82.36         & 1753       & 1972       \\
G\^{a}rlea~\cite{Garlea1978Measuring}& 2325$\pm$93& 1512$\pm$53& 83.46$\pm$3.3 & 1780$\pm$62& 2026$\pm$81\\
\hline
\end{tabular}
\end{center}
\end{table}

\subsection{Validation by small-sized LANL fast system criticalities}
Figure~\ref{fig:diffJ40} shows the $^{235}$U and $^{239}$Pu cross sections from the present evaluation are higher than the JENDL-4.0 cross sections in the 13rd to 17th groups (143--498~keV).
It is known that the criticalities of the small-sized LANL fast systems such as Godiva and Jezebel are sensitive to the $^{235}$U and $^{239}$Pu cross sections in this energy region,
and one can expect the criticalities calculated with the newly evaluated cross sections are much higher than those calculated with the JENDL-4.0 cross sections unless other data (e.g., capture cross sections, prompt fission neutron multiplicities and spectra) are readjusted.
As the cross sections of this energy region are important for fast reactor application,
the criticalities of nine small-sized LANL fast systems (Jezebel-23, Flattop-23, Godiva, Flattop, Big-Ten, Jezebel, Jezebel-240, Flattop-Pu and Thor) were calculated by Yasunobu Nagaya (JAEA) with the original JENDL-4.0 library and its update prepared by DeCE.
The neutron transport calculations were performed with the Japanese continuous-energy Monte Carlo code MVP Version 3~\cite{Nagaya2017MVP}.

\begin{figure}[hbtp]
\centering\includegraphics[clip,angle=-90,width=0.8\linewidth]{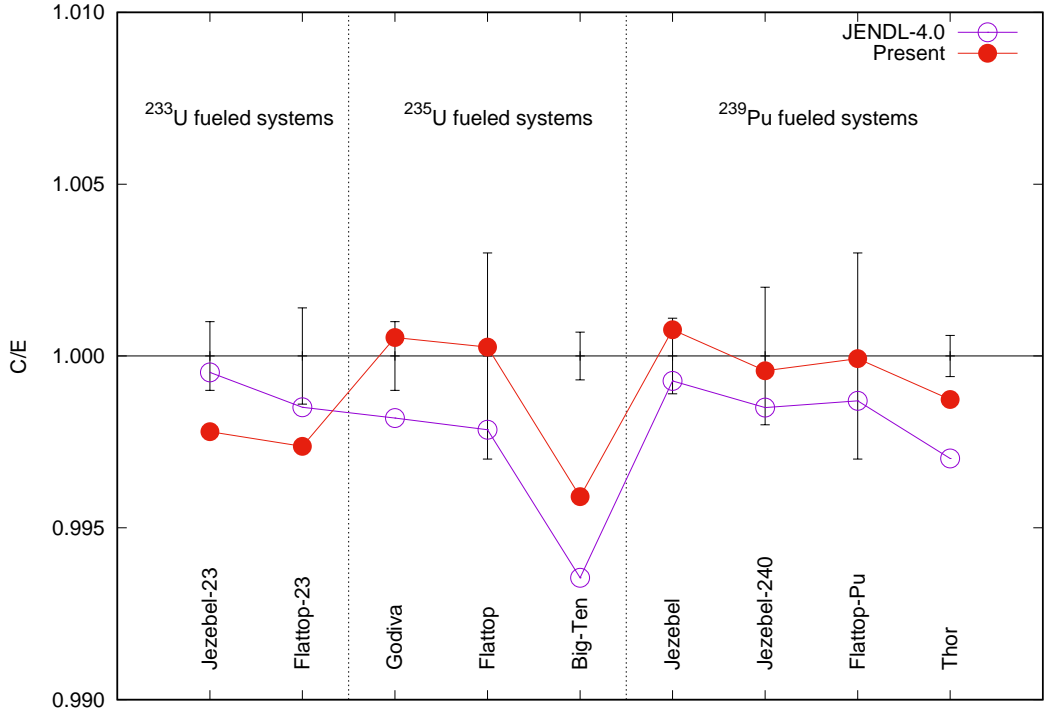}
\caption{C/E values of the LANL small-sized fast system criticalities calculated by MVP Version 3 with the cross sections in JENDL-4.0 and those updated with the present evaluation.
The standard deviation in the calculated criticality is within the size of the symbol.
Courtesy of Yasunobu Nagaya (JAEA).
}
\label{fig:keff}
\end{figure}

Figure~\ref{fig:keff} compares the C/E values of the criticalities calculated with the original and updated JENDL-4.0 cross sections.
The changes in the C/E values of the criticalities due to update from JENDL-4.0 are similar to those seen in the Cf-252 SACS.
Namely the cross section update leads to decrease in the C/E values for the $^{233}$U fueled systems while to increase of the C/E values for the $^{235}$U and $^{239}$Pu fueled systems.
The tendencies for the $^{235}$U and $^{239}$Pu fueled systems are also understandable from the increase of their cross sections in the 150--450~keV region seen in Fig.~\ref{fig:diffJ40}.
The C/E value of Big-Ten is improved with the current evaluation but it is still considerably low.
The JENDL-4.0 benchmark summary~\cite{Chiba2011JENDL-4.0} mentions the Big-Ten C/E value is sensitive to $^{238}$U inelastic scattering cross section.
But the fission cross section of our evaluation lower than 100~keV could be also partly responsible to the low Big-Ten C/E value since the Big-Ten neutron spectrum is softer than the spectra of the other $^{235}$U fueled systems.
Figure~\ref{fig:diffI17} shows our $^{235}$U cross section is systematically lower than IAEA-2017 ($\sim$ENDF/B-VIII.0) below 100~keV.

We also analyzed the low C/E values for the Jezebel-233 and Flattop-233 by using the sensitivities of the cross sections with respect to the criticalities calculated by a deterministic reactor physics code CBZ~\cite{Chiba2016Nuclear}.
It indicates that the decrease in their criticalities is due to update of the $^{233}$U cross section around 2~MeV from JENDL-4.0.

\subsection{High energy cross sections}
Some JENDL special purpose libraries include the $^{235,238}$U and $^{239,240,241}$Pu fission cross sections above 20 MeV evaluated by an approach completely different from the present approach.
This time we evaluated the fission cross sections of the six nuclides above 20 MeV for the JENDL general purpose library for the first time.
The prior $^{235,238}$U and $^{239,240,241}$Pu cross sections taken from the JENDL-4.0/HE library are those calculated by the GNASH code~\cite{Young1998Comprehensive} and slightly changed to fit to the available experimental datasets according to the descriptions in their ENDF-6 files.
The prior $^{233}$U cross section evaluated by Yavshits et al. is theoretical one obtained by their multiconfiguration fission approach.
Most of the high energy experimental datasets adopted in the present evaluation were not available when these prior cross sections were evaluated.
Here we discuss comparison of the high energy cross sections from the present evaluation with their prior cross sections as well as the cross sections from time-of-flight measurements.

\subsubsection{$^{233}$U}
Figures~\ref{fig:U233log} and \ref{fig:U233U235log} show that the newly evaluated $^{233}$U high energy cross section is systematically lower than the prior cross section evaluated by Yavshits et al.
The two experimental $^{233}$U/$^{235}$U cross section ratios measured by Tovesson et al. and Shcherbakov et al. are consistent each other in the high energy region.
The number of target atoms based on sample specifications is used by Tovesson et al. while the ratio measured by Shcherbakov et al. is normalized to the JENDL-3.2 cross section ratio at 1.75--4.0~MeV.

\subsubsection{$^{235}$U}
Figure~\ref{fig:U235log} shows that the newly evaluated high energy cross section is very close to the JENDL-4.0/HE cross section.
Probably the JENDL-4.0/HE cross section largely relies on the cross section measured by Lisowski et al. though its use is not described in the JENDL-4.0/HE data file.
The newly evaluated high energy absolute cross sections of not only $^{235}$U but all nuclides strongly depend on this Lisowski et al's measurement.
For example, we observe influence of the structure around 50~MeV in this $^{235}$U measurement on the newly evaluated cross sections of other nuclides (e.g., Figs.~\ref{fig:Pu239log} and \ref{fig:Pu240log}).
The high energy cross section should be reviewed again when the $^{235}$U cross section from 10~MeV to 1 GeV measured relative to n-p scattering at the CERN n\_TOF facility~\cite{Manna2020Setup} is published and added to the EXFOR library.

\subsubsection{$^{238}$U}
Figures~\ref{fig:U238log} and \ref{fig:U238U235log} show that the newly evaluated $^{238}$U high energy cross section is close to the JENDL-4.0/HE cross section.
Its data file describes that the evaluation considered three measurements~\cite{Lisowski1997,Donets1999Neutron, Eismont1996Relative}, among which the second and third ones are most probably from the measurements by Lisowski et al. and Shcherbakov et al adopted by us.
Major differences are seen among the high energy experimental ratios, and we plotted in Fig.~\ref{fig:exp-sok-rat} these experimental and prior (JENDL-4.0 and JENDL-4.0/HE) ratios relative to the ratio from the present evaluation.
The JENDL-4.0/HE ratio is between the two ratios measured by Lisowski et al. and Shcherbakov et al., and the situation is similar in the present evaluation.
Among the experimental ratios newly considered in the present evaluation, the fission ionization chamber (FIC) dataset measured by Paradela et al. (EXFOR 23269.002) is close to the present evaluation while the present evaluation is close to or even below the lower boundaries of the error bars of the other new ratios measured by Paradela et al. and Tovesson et al.
Note that Shcherbakov et al. use the threshold method~\cite{Behrens1981Measurement} instead of sample quantification for overall normalization of their ratio.

\subsubsection{$^{239}$Pu}
\begin{figure}[hbtp]
\centering\includegraphics[clip,width=0.8\linewidth]{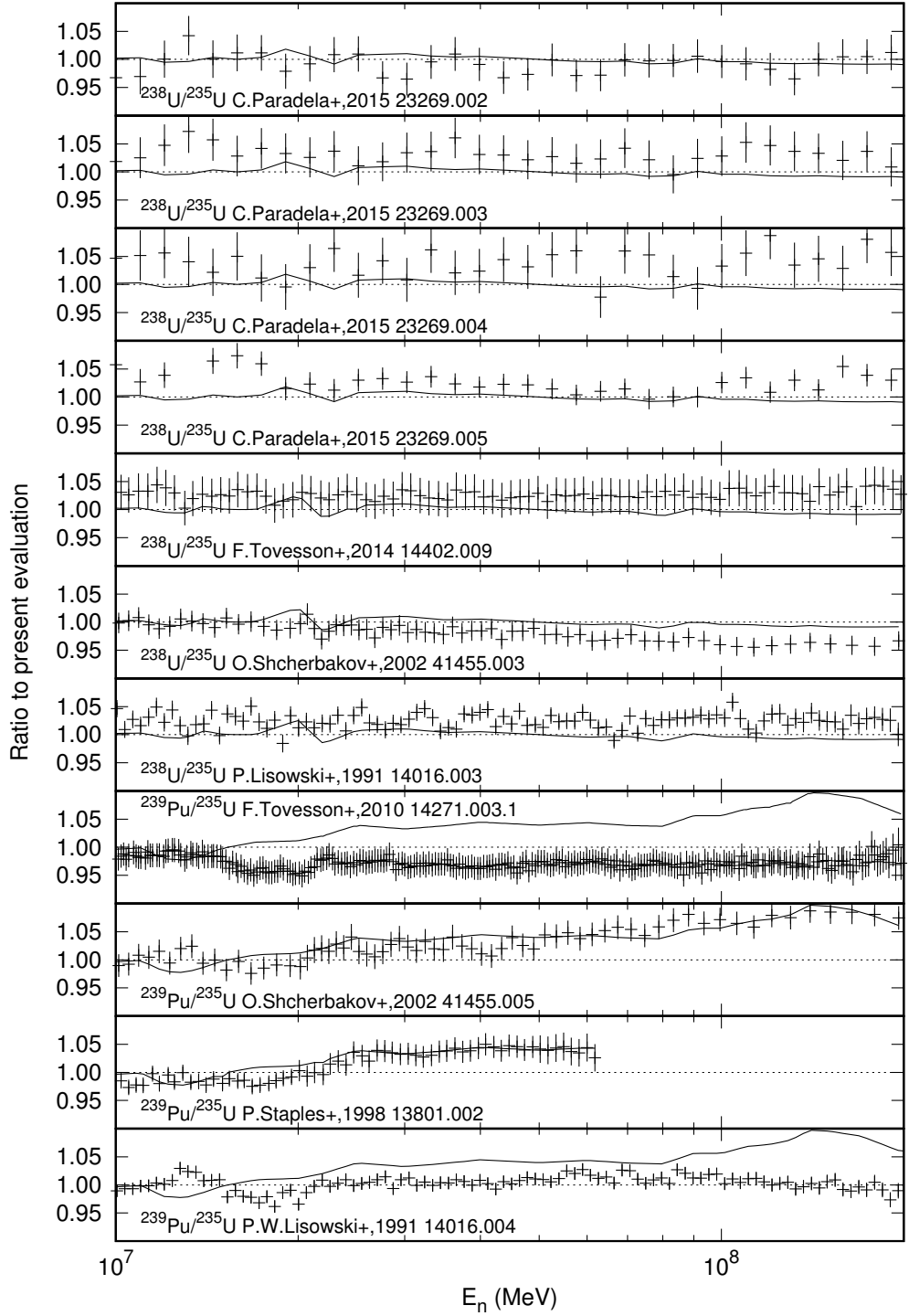}
\caption{
	The $^{238}$U/$^{235}$U and $^{239}$Pu/$^{235}$U cross section ratios from measurements (symbol) and prior (JENDL-4.0 and JENDL-4.0/HE) cross sections (solid line) relative to the ratios from the present evaluation.
The EXFOR 23269.003 and 23269.004 datasets by Paradela et al. are from two configurations of the same sample and compiled together in our experimental database as a single dataset (EXFOR 51005.002).
}
\label{fig:exp-sok-rat}
\end{figure}

Figures~\ref{fig:Pu239log} and \ref{fig:Pu239U235log} show that the newly evaluated $^{239}$Pu high energy cross section is significantly lower than the JENDL-4.0/HE cross section.
Unlike the $^{238}$U/$^{235}$U case, we observe the JENDL-4.0/HE ratio is largely influenced by the ratio measured by Shcherbakov et al. and not by Lisowski et al.
The ratio from the present evaluation agrees with the ratio measured by Lisowski et al. which is most probably not considered in the JENDL-4.0/HE evaluation.
Figure~\ref{fig:exp-sok-rat} shows that the ratio from the present evaluation is close to the upper boundary of the error bars of the ratio measured by Tovesson et al. in general, but systematically lower than the ratio measured by Shcherbakov et al. and Staple et al.
The ratios measured by Tovesson et al. and Shcherbakov et al. are normalized to the thermal cross section ratio and the JENDL-3.2 library at 1.75--4.0~MeV, respectively, while Staple et al. quantify the target atoms by alpha spectrometry.
The IAEA-2017 evaluation summary~\cite{Carlson2018Evaluation} mentions a preliminary result of a very accurate fission cross section ratio measurement by a time projection chamber of the NIFFTE collaboration (not in EXFOR) is in excellent agreement with IAEA-2006, and we expect the newly measured ratio is closer to the ratio measured by Lisowski rather than the ratios measured by Shcherbakov et al. and Tovesson et al. (See Fig.~37(c) of Ref.~\cite{Carlson2018Evaluation}).

\subsubsection{$^{240}$Pu}
Figures~\ref{fig:Pu240log} and \ref{fig:Pu240U235log} show that the newly evaluated $^{240}$Pu high energy cross section is close to the JENDL-4.0/HE cross section in general.
The ratios measured by Laptev et al. and Staples et al. are consistent with the new evaluation.
The ratio measured by Laptev et al. is normalized to IAEA-2006 at 1, 5 and 10~MeV while Staple et al. quantify the target atoms by alpha spectrometry.

\subsubsection{$^{241}$Pu}
Figures~\ref{fig:Pu241log} and \ref{fig:Pu241U235log} show the cross section from the present evaluation is systematically lower than the JENDL-4.0/HE evaluation.
We observe in Fig.~\ref{fig:Pu241U235log} that the current evaluation adopts the shape of the ratio measured by Tovesson et al. at the high energy region but its overall normalization is more consistent with the ratio measured by Fursov et al. between 1 and 10~MeV.

\subsection{Uncertainty and correlation}
\label{sec:unc}
\begin{figure}[hbtp]
\centering\includegraphics[clip,angle=-90,width=0.8\linewidth]{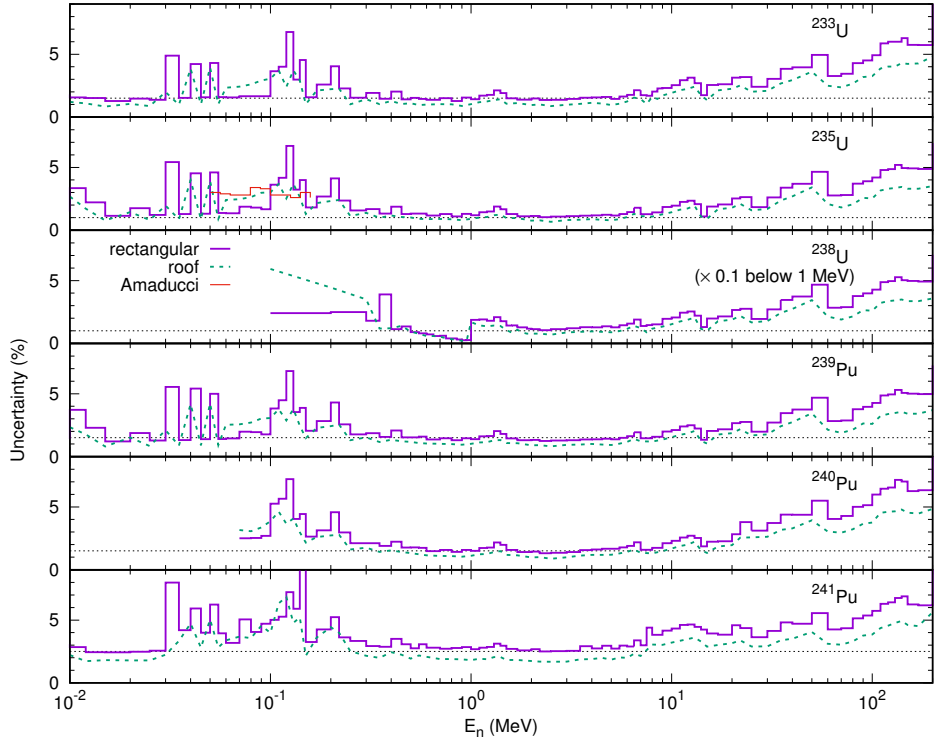}
\caption{
External uncertainties in the fitting parameters in the roof function and rectangular function expressions.
The uncertainty in the parameter in the rectangular function expression is equal to the uncertainty in the group-wise cross section.
The horizontal dashed line indicates best achievable precision for each nuclide.
The additional histogram (thin solid line) between 50~keV and 160~keV of the $^{235}$U panel shows the uncertainty in the cross section measured by Amaducci et al.~\cite{Amaducci2019Measurement}.
}
\label{fig:unc}
\end{figure}

Figure~\ref{fig:unc} shows the uncertainties in the fitting parameters $\{\sigma_j\}$ in the roof function expression. 
The uncertainty in the evaluated cross section at any energy may be obtained by propagation from the uncertainties in $\{\sigma_j\}$ according to the model function.
Alternatively,
the uncertainty in the group-wise cross section at the energy interval $[E_j, E_{j+1}]$ may be obtained by fitting with the rectangular function
\begin{equation}
\Delta_j(E)=
\left\{
\begin{array}{ll}
1 & (E_j \le E \ < E_{j+1})  \\
0 & \textrm{otherwise}       \\
\end{array}
\right.
,
\label{eqn:rectangular}
\end{equation}
and we evaluated the uncertainty in the cross section from the present evaluation by this approach since the covariance of the group-wise cross section can be accommodated in the ENDF-6 format.
By fitting with the rectangular function,
we obtained the fitting parameters (= group-wise cross sections) with the reduced chi-square of 10.3.
The reduced chi-square with the roof function expression is smaller (4.00),
and it indicates the roof function is more adequate for modelling of the energy dependence of the cross sections.
The uncertainty in the group-wise cross section from the rectangular function expression is shown in Fig.~\ref{fig:unc}.
The figure shows the best precision of about 1\% ($^{235,238}$U), 1.5\% ($^{233}$U, $^{239,240}$Pu) or 2.5\% ($^{241}$Pu) is achieved around 2 to 3~MeV in the rectangular expression.
The figure also indicates presence of the large uncertainties for all nuclides in 50-250~keV in the roof function expression,
where the uncertainty in our $^{235}$U cross section is close to the uncertainty reported by the most recent measurement by Amaducci et al.~\cite{Amaducci2019Measurement}.
Few $^{235}$U experimental datasets were usable for our analysis in this energy region,
and our evaluation could be improved for all six target nuclides if additional well-documented new $^{235}$U cross sections become available in this energy region.

Note that the reduced chi-square higher than 1 indicates the uncertainty from the least-squares fitting (internal uncertainty) underestimates the uncertainty expected from the actual discrepancy of the experimental datasets (external uncertainty),
and we converted the internal uncertainty to the external uncertainty by multiplying the internal uncertainty by the square root of the reduced chi-square.
All uncertainties plotted in Fig.~\ref{fig:unc} are the external ones.
We are aware that a shortcoming of this approach is treatment of all experimental data points of the various reactions and energies equally~\cite{Capote2020Unrecognized}.
The JENDL-3.3 evaluators consider the uncertainty obtained by their evaluation is too small probably due to unknown systematic uncertainties of the measurements~\cite{Kawano2000Simultaneous}.
But the large chi-square value may be also originated from a bias effect (missing correction).
``{\it Unknown influences on a measurement result can, obviously, not be taken into consideration and can therefore not be included in an uncertainty. ... Not correcting does not cause a systematic uncertainty but a systematic deviation, i.e., just a wrong result}"~\cite{Drosg2007Dealing}.
We consider that high quality evaluation can be achieved by use of experimental datasets which pay attention not only to higher precision (e.g., by improving uncertainties) but also to higher accuracy (e.g., by identifying overlooked bias effects)~\cite{Harada2021Dealing}.

\begin{figure}[hbtp]
\centering\includegraphics[clip,angle=-90,width=0.8\linewidth]{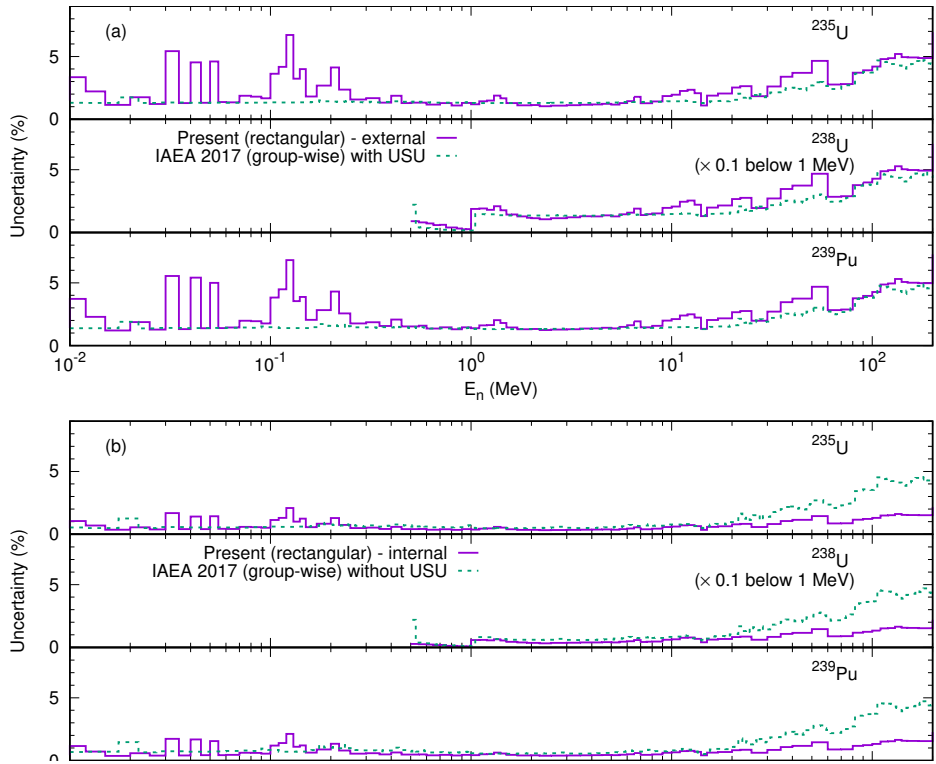}
\caption{
Uncertainties from the present evaluation (rectangular function expression) and IAEA Neutron Data Standards 2017 evaluation (group-wise cross section).
(a) External uncertainties from the present evaluation and uncertainties including the unrecognized systematic uncertainties (USU) from the IAEA-2017 evaluation.
(b) Internal uncertainties from the present evaluation and uncertainties excluding USU from the IAEA-2017 evaluation.
}
\label{fig:unc-i17j5a}
\end{figure}

Figure~\ref{fig:unc-i17j5a} compares the uncertainties from the present evaluation in the rectangular function expression with those in IAEA-2017.
The latter uncertainties are for the group-wise cross sections compiled in the ENDF-6 format.
The upper part (a) of the figure is for the final uncertainties of the evaluations, namely the external uncertainties for the present evaluation and the uncertainties including the unrecognized systematic uncertainty (USU) for the IAEA evaluation.
The lower part (b) of the figure is for the internal uncertainties for the present evaluation and the uncertainties excluding the USU for the IAEA evaluation.
The USU is estimated to be 1.2\%~\cite{Carlson2018Evaluation} for all three nuclides and its influence on the uncertainty in the IAEA-2017 cross section is small in the figure.
The IAEA evaluation introduces in the GMA database an additional uncertainty component to outlying experimental data and it brings a reduced chi-square closer to 1, and it is included in the IAEA cross section uncertainties in the upper and lower part of the figure.
Two evaluations show similar final precision except for 100--200~keV region and a few groups in the 30--60~keV region,
where the uncertainty from the present evaluation shows strong energy dependence but it is rather constant in the IAEA evaluation.
The lower part of the figure shows that the uncertainty from the present evaluation is slightly lower than the IAEA evaluation if we do not enlarge the uncertainty estimated by the SOK by the square root of the reduced chi-square.
Note that the groups of the present evaluation are wider than those of the IAEA evaluation (the number of groups between 10~keV and 200~MeV is about 80 and 140 in the present and IAEA evaluation, respectively),
and therefore the comparison in the uncertainties on this figure must be done with caution.

\begin{figure}[hbtp]
\centering\includegraphics[clip,angle=-90,width=0.8\linewidth]{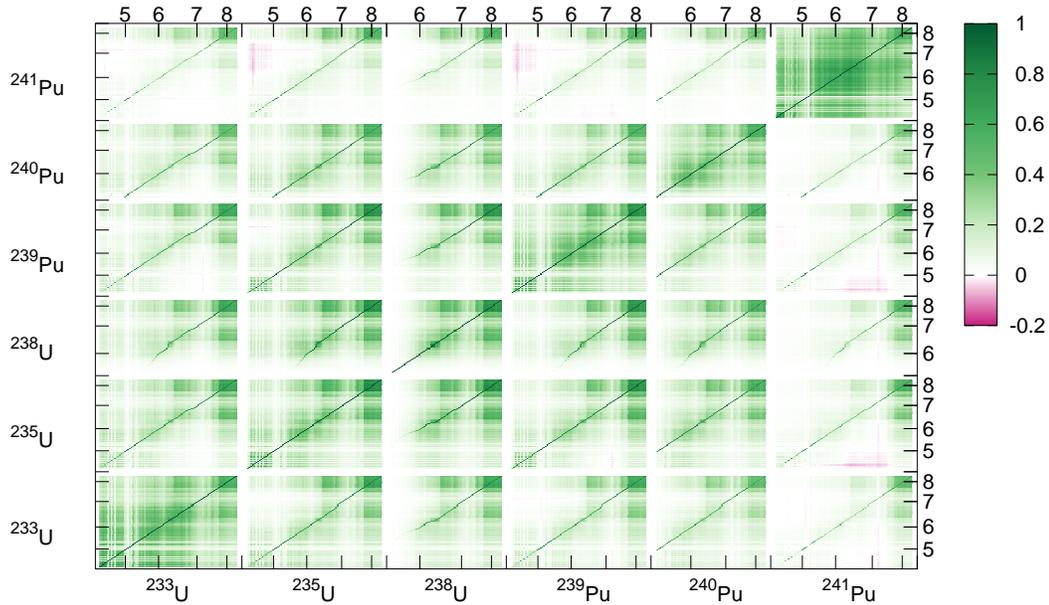}
\caption{
Correlation coefficients of the fitting parameters from the present evaluation in the rectangular function expressions.
The numbers on the ticks along the axes on the upper and right sides give the exponentials of the incident energies in eV by power to base 10 (e.g., 6 for $10^6$ eV).
}
\label{fig:cor}
\end{figure}

Figure~\ref{fig:cor} shows the correlation coefficients between the evaluated cross sections (fitting parameters) between the energy groups and target nuclides in the rectangular expression.
We observe presence of the strong correlation between any two target nuclides at the same incident energy except for $^{238}$U in the subthreshold fission region (below $\sim$1~MeV).

\section{Summary}
The fission cross sections of $^{233,235,238}$U and $^{239,240,241}$Pu were evaluated by using the simultaneous least-squares fitting code SOK for the JENDL-5 library.
The evaluated cross sections were obtained in the energy range of 10~keV ($^{233,235}$U, $^{239,241}$Pu) or 100~keV ($^{238}$U, $^{240}$Pu) to 200~MeV.
The cross sections in the high energy region (20--200~MeV) were evaluated for the JENDL general purpose library for the first time.
Each EXFOR entry relevant to our evaluation was reviewed against the source article,
especially for the uncertainty information,
and the EXFOR entry was updated by the originating Data Centre when necessary.
Also special attention was paid during construction of our experimental database to exclude double counting of the information from the same experimental work.
The experimental datasets compiled in the EXFOR entries with the information sufficient for covariance matrix construction were converted to the input format of SOK with a few manipulations such as elimination of some data points and addition of normalization uncertainty inferred from the source article.

The outputs from the least-squares fitting code SOK were adopted as the evaluated cross sections without any further corrections.
The changes in the obtained evaluated cross sections from those in the JENDL-4.0 library are within 4\% ($^{241}$Pu), 3\% ($^{233}$U, $^{240}$Pu), or 2\% ($^{235}$U, $^{239}$Pu).
The best precisions of the group-wise cross sections were achieved around 2 to 3~MeV,
where the external uncertainties are 1\% ($^{235,238}$U), 1.5\% ($^{233}$U, $^{239}$Pu) or 2.5\% ($^{241}$Pu).

The californium-252 spectrum averaged cross sections were calculated from the evaluated cross sections,
and compared with those measured by Grundl et al. and recommended by Mannhart. 
The comparison shows reasonable agreement except for $^{238}$U,
for which our new evaluation could underestimate the actual cross section in the energy region around 2~MeV.
The newly evaluated cross sections averaged over the $\Sigma\Sigma$ neutron spectrum agree with those measured by G\^{a}rlea et al. for the $^{233,235,238}$U and $^{239}$Pu cross sections especially when the comparison is done for the ratios to the $^{235}$U cross section.
A benchmark calculation performed for small-sized LANL fast system criticalities by MVP shows a reasonable agreement except for two $^{233}$U fueled systems and Big-Ten.

The final result of the present evaluation (SOK 20210404) was submitted for preparation of the JENDL-5 library.
The JENDL-5 library would become the most up-to-date data library in terms of our experimental knowledge of the fission cross sections of the six target nuclides in the fast neutron region.
Not only the JENDL project but also other data library projects may benefit from the uncertainty information added from the source articles to the EXFOR entries during the present evaluation.

Improvement of the $^{240}$Pu fission cross section between 0.5 and 5~MeV and the $^{241}$Pu fission cross section between 9~keV and 1~MeV are required to meet the target accuracy for the core and fuel cycle of a wide range of innovative systems~\cite{Salvatores2008Role,Salvatores2008Uncertainty}, and their target accuracies in the energy region of the present evaluation are 2 to 3\% for fast reactor systems such as sodium-cooled fast reactor (SFR), gas-cooled fast reactor (GFR) and lead-cooled fast reactor (LFR) according to the NEA High Priority Request List (HPRL)~\cite{Dupont2020HPRL}.
The newly obtained evaluation cross sections do not satisfy this requirement,
and further measurements and reevaluation are necessary.
The present evaluation also shows relatively high uncertainty in 100-300~keV,
and an additional measurement of the $^{235}$U cross section in this region would improve the precision of the newly evaluated cross sections not only for $^{235}$U but also for other target nuclides.

\section*{Acknowledgement}
The preceding simultaneous evaluation for the JENDL-4.0 library was performed under excellent supervision of Tsuneo Nakagawa (Japan Atomic Energy Agency),
which gave us strong motivation to perform the present evaluation.
The authors are grateful to Toshihiko Kawano (Los Alamos National Laboratory) for allowing us use of the SOK code and also providing valuable comments on its outputs.

Vladimir Pronyaev (Institute of Physics and Power Engineering) and Allan Carlson (National Institute for Standards and Technology) helped us to understand the status of the experimental data compiled in the GMA database.
The numerical $\Sigma\Sigma$ spectrum in the 725 energy group structure was received from Stanislav Simakov (Karlsruhe Institute of Technology).
Knut Merla, Guntram Pausch and Friedrich Vo\ss  helped us to resolve the questions on the relations of similar datasets compiled in EXFOR for those measured in Dresden and Karlsruhe many decades ago.

The Monte Carlo benchmark calculation for the LANL small-sized fast system criticalities was performed by Yasunobu Nagaya (Japan Atomic Energy Agency).
Part of the uncertainty information of the $^{238}$U and $^{240}$Pu fission cross sections in the EXFOR library was reviewed against the source articles with Amanda Lewis (IAEA intern from University of California, Berkeley).
The NDS EXFOR and ENDF web retrieval and plotting systems maintained and developed by Viktor Zerkin (International Atomic Energy Agency) supported this work.

We received valuable comments on the manuscript from anonymous referees and Hideo Harada (Japan Atomic Energy Agency).
We also thank Go Chiba (Hokkaido University), Keiichi Shibata (Japan Atomic Energy Agency) and Roberto Capote (International Atomic Energy Agency) for their interests and supports to this work.

Last but not least, we would like to express our appreciation to all experimentalists who submitted their numerical data to the EXFOR library and responded to our questions with patience, and also to the past and present EXFOR compilers and managers of the Nuclear Reaction Data Centres (NRDC) for their continuous dedication and support to the internationally maintained EXFOR library.
\bibliography{jendl5-arxiv}

\end{document}